\documentclass[iop]{emulateapj}

\shorttitle{Variability in 55 Cnc \textit{e}'s Eclipse Depth}
\shortauthors{Tamburo et al.}

\begin{document}


\title{Confirming Variability in the Secondary\\
    Eclipse Depth of the Super-Earth 55 Cancri e}


\author{P. Tamburo\altaffilmark{1,2,3}, A. Mandell\altaffilmark{2}, D. Deming\altaffilmark{3,4}, and E. Garhart\altaffilmark{5}} 

\affil{
\altaffilmark{1}Boston University, 1 Silber Way, Boston, MA 02215\\
\altaffilmark{2}NASA Goddard Space Flight Center, 8800 Greenbelt Rd., Greenbelt, MD 20771\\
\altaffilmark{3}University of Maryland College Park, College Park, MD 20742\\
\altaffilmark{4}NASA Astrobiology Institute's Virtual Planetary Laboratory\\
\altaffilmark{5}Arizona State University, Tempe, AZ 85281}
\email{tamburop@bu.edu}




\begin{abstract}
We present a reanalysis of five transit and eight eclipse observations of the ultra-short period super-Earth 55 Cancri \textit{e} observed using the \textit{Spitzer Space Telescope} during 2011-2013. We use pixel-level decorrelation to derive accurate transit and eclipse depths from the \textit{Spitzer} data, and we perform an extensive error analysis. We focus on determining possible variability in the eclipse data, as was reported in \citet{Demory2016a}. From the transit data, we determine updated orbital parameters, yielding T0 = 2455733.0037 $\pm$ 0.0002, P = 0.7365454 $\pm$ 0.0000003 days, i = 83.5 $\pm$ 1.3 degrees, and $R_p$ = 1.89 $\pm$ 0.05 $R_\oplus$. Our transit results are consistent with a constant depth, and we conclude that they are not variable. We find a significant amount of variability between the eight eclipse observations, and confirm agreement with \citet{Demory2016a} through a correlation analysis. We convert the eclipse measurements to brightness temperatures, and generate and discuss several heuristic models that explain the evolution of the planet's eclipse depth versus time. The eclipses are best modeled by a year-to-year variability model, but variability on shorter timescales cannot be ruled out. The derived range of brightness temperatures can be achieved by a dark planet with inefficient heat redistribution intermittently covered over a large fraction of the sub-stellar hemisphere by reflective grains, possibly indicating volcanic activity or cloud variability. This time-variable system should be observable with future space missions, both planned (\textit{JWST}) and proposed (i.e. \textit{ARIEL}).
\end{abstract}





\section{Introduction}

Planets with extremely short orbital periods (P $<$ 0.75 days) are subject to intense irradiation from their host stars, which causes high surface temperatures and can drive significant atmospheric loss rates \citep[e.g.][]{Valencia2010}. In addition, they experience tidal forces that potentially produce a dissipation of energy that can increase their temperatures by several degrees per year if they fail to achieve a 1:1 spin-orbit resonance \citep[][]{Makarov2014}. The formation history, orbital evolution and compositional structure of these planets are poorly understood, and they therefore represent a significant challenge in furthering our understanding of the variety of planetary systems.

One of the most well-studied ultra-short period planets is 55 Cancri \textit{e}, a transiting super-Earth orbiting a Sun-like star about 13 parsecs from Earth. Detected by \citet{McArthur2004} and confirmed by \citet{Fischer2008}, the planet was originally thought to have a 2.8 day period and a minimum mass of 14 ${M_{\earth}}$. \citet{Dawson2010}, however, proposed that the detected period and mass were due to aliasing in the radial velocity data, and that the planet's true period and mass were 0.74 days and 8 ${M_{\earth}}$. \citet{Winn2011} and \citet{Demory2011} made photometric observations that confirmed the proposed values.

A rapid orbit around a bright star that permits relatively high signal-to-noise observations \citep[][]{von Braun2011} makes 55 Cnc \textit{e} a favorable target for performing frequent transit and eclipse observations that allow for further characterization of the planet. Our desire to understand the nature of super-Earths has resulted in a plethora of data for the system \citep[e.g.][]{Dragomir2014,Demory2016a}, and various studies have attempted to characterize the planet's structure and composition. \citet{Demory2016b}, for example, created a day-night temperature map of the planet using phase-curve data, while \citet{Lopez2017} found that the planet must have a water envelope of $8 \pm 3\%$ of the total planet mass to be consistent with the mass and radius.

One of the most notable results is that of \citet{Demory2016a} (hereafter D16), who found that the secondary eclipse depth of the planet increased significantly over the course of eight observations in the 4.5\,$\mu$m channel of the \textit{Spitzer Space Telescope} \citep[][]{Werner2004} between 2012 and 2013. \textit{Spitzer} has yielded high precision photometry that has allowed for the characterization of many diverse exoplanetary systems \citep[e.g.][]{Gillon2017,Knutson2012}. The increase in eclipse depth is indicative of an increase of the planet's brightness temperature ($T_B$) at 4.5\,$\mu$m from 2012 to 2013, and D16 found that $T_B$ increased from about 1600K to 2600K over that time span. D16 posited that the variation could possibly be accounted for by volcanism on the planet. Volcanic plumes could in principle raise the photosphere at 4.5\,$\mu$m higher in the atmosphere where the temperature is lower, leading to lower thermal emission when the plumes are active \citep[][]{Demory2016a}.

Genuine detection of eclipse variability would therefore be an extraordinary result, as it could indicate a volcanic process occurring on a body outside of our solar system. Because extraordinary results require extraordinary evidence, we present here a reassessment of the possible variability in eclipse and transit depths for this system. We not only use a new and powerful method to correct for the instrumental signatures in the data, but we also develop a new method for error analysis that provides an independent check on our quoted errors. Moreover, we evaluate a wider range of heuristic models for eclipse variability than did D16. In Section \ref{sec:DataAnalysis}, we describe the observations of 55 Cnc \textit{e} that were analyzed, as well as our data analysis procedures. In Section \ref{sec:Results}, we discuss the results for both the transit and eclipse observations. Section \ref{sec:Discussion} discusses the implications of our analysis, and how the implications of our results relate to the thermal variability of 55 Cnc \textit{e}.

\section{Data Analysis}
\label{sec:DataAnalysis}
\subsection{Observations}
\label{sec:Observations}
	We examined eight eclipses and six transits of 55 Cnc \textit{e} taken in the 4.5\,$\mu$m channel of the \textit{Spitzer Space Telescope's} IRAC instrument \citep[][]{Fazio2004} using subarray mode. These data are identical to those analyzed by D16, and are publicly available through the Spitzer Heritage Archive (http://sha.ipac.caltech.edu/applications/Spitzer/SHA/). We list the date, type, program ID, and AOR of each observation in Table \ref{tab:data_table}.
	
    All observations were taken with 0.02-s exposures using subarray mode. Observations taken after the 2012 January 18 eclipse were all taken with PCRS peak-up, which is designed to limit stellar motion on the detector and hence reduce the intra-pixel effect \citep{Ingalls2012}. Table 1 gives the date, observation type, and AOR number for all the data used.
	
\begin{table*}
        \centering
        \caption{Date, type, program ID, and AOR number of observations analyzed in this work. We also list the percentage of frames discarded ($Fr_{disc}$) and pixels corrected ($Pix_{corr}$) for each observation, per our discussion in Section \ref{sec:photometry}. $\sigma_{phot}$ are the photon-limited errors of the unbinned data, while $\frac{SDNR}{\sigma_{phot}}$ is the ratio of the standard deviation of unbinned residuals (photometry-fit) to the photon-limited errors. On average, the scatter of our residuals comes within 10\% of the photon noise limit.}
        \label{tab:data_table}
        \begin{tabular}{cccccccc}
            \hline
            Date & Type & Program ID & AOR & $Fr_{disc}$ ($\%$) & $Pix_{corr}$ ($\%$) & $\sigma_{phot}$ (ppm) & $\frac{SDNR}{\sigma_{phot}}$\\
            \hline
            2011 Jan 6 & Transit & 60027 & 39524608 & 0.034 & 0.0006 & 7437 & 1.08\\
            2011 Jun 20 & Transit & 60027 & 42000384 & 0.010 & 0.0004 & 7414 & 1.07\\
            2012 Jan 18 & Eclipse & 80231 & 43981056 & 0.001 & 0.0020 & 7511 & 1.27\\
            2012 Jan 21 & Eclipse & 80231 & 43981312 & 0.042 & 0.0007 & 7418 & 1.11\\
            2012 Jan 23 & Eclipse & 80231 & 43981568 & 0.044 & 0.0005 & 7473 & 1.09\\
            2012 Jan 31 & Eclipse & 80231 & 43981824 & 0.093 & 0.0005 & 7490 & 1.14\\
            2013 Jun 15 & Eclipse & 90208 & 48070144 & 0.084 & 0.0020 & 7374 & 1.03\\
            2013 Jun 18 & Eclipse & 90208 & 48073216 & 0.047 & 0.0008 & 7426 & 1.07\\
            2013 Jun 21 & Transit & 90208 & 48070656 & 0.053 & 0.0023 & 7397 & 1.10\\
            2013 Jun 29 & Eclipse & 90208 & 48073472 & 0.036 & 0.0006 & 7437 & 1.05\\
            2013 Jul 3 & Transit & 90208 & 48072448 & 0.050 & 0.0016 & 7428 & 1.09\\
            2013 Jul 8 & Transit & 90208 & 48072704 & 0.044 & 0.0036 & 7418 & 1.14\\
            2013 Jul 11 & Transit & 90208 & 48072960 & - & - &  -   & - \\
            2013 Jul 15 & Eclipse & 90208 & 48073728 & 0.026 & 0.0006 & 7444 & 1.11\\
            \hline
        \end{tabular}
\end{table*}

Intra-pixel sensitivity fluctuations in the 4.5\,$\mu$m channel can cause flux variations of about 1\%, depending on how the stellar PRF moves over the course of observations \citep[e.g.][]{Charbonneau2005, Mighell2008}. This effect is two orders of magnitude larger than the expected eclipse depths, and has to be accurately removed in order to draw scientific conclusions from the observations. Our process for removing the intrapixel fluctuations is detailed in Section \ref{sec:Fitting for Eclipse Depths}.

\subsection{Photometry}
\label{sec:photometry}
The data comprise cubes of 64 frames, each having 32x32 pixels. The star is very bright and our exposures very short, so background is negligible. Nevertheless, we determine the average background value by first fitting a Gaussian to a histogram of pixel intensities over the entire 32x32 pixel image to determine the centroid value. This value was then subtracted from our images.

Because knowing the position of the star in the images is necessary in order to place a numerical aperture and perform accurate photometry, we found the location of the star in the images using two different methods. First, we determined the centroid of the stellar image using a 2-D Gaussian fitting procedure in IDL. Second, we performed a center-of-light (COL) calculation, that separately defined the centroid of the stellar image in each coordinate. We collapsed the image in the Y dimension, then calculated the centroid in X by intensity weighting the X-coordinates, and vice-versa to calculate the image centroid in Y. Both of these fits were performed using the full 32x32 pixel images. Having determined the center of the star, we performed aperture photometry for both the 2-D Gaussian and COL methods.

For both centering methods, we summed the stellar flux in 11 numerical circular apertures of various radii, ranging from 1.6 to 3.5 pixels. We used both constant and time-variable radii, the latter being determined using the ``noise pixel" implementation of \citet{Lewis2013}, to which we added constant increments in radius ranging from 0.1 to 2 pixels. The two approaches for identifying the location of the star along with the option for fixed and time-variable apertures gave us four different photometry files that can be used in our fitting code.

\begin{figure}
    \centering
    \includegraphics[scale=0.42]{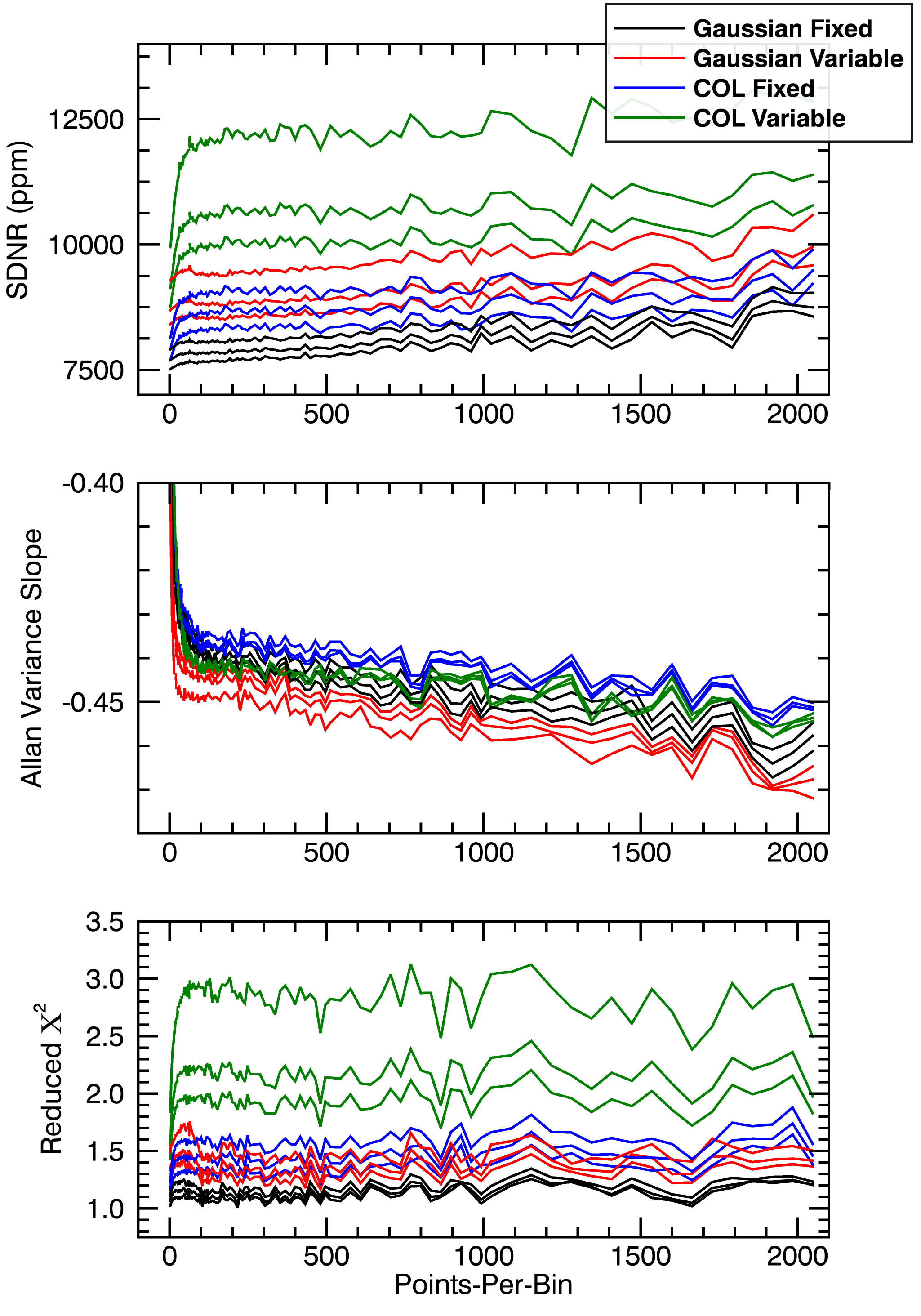}
    \caption{Standard deviation of normalized residuals, Allan variance slope, and reduced $\chi^2$ versus bin size for our four different photometry methods for the 2012 January 31 eclipse. Similar results for other observations motivated us to limit ourselves to using the photometry with Gaussian centroiding and fixed apertures exclusively. The different strands of similarly-colored lines correspond to different aperture sizes used to perform the photometry.}
    \label{fig:phot_method_comp}
\end{figure}

In practice, we found that it is advantageous to limit ourselves to using only the photometry files that use Gaussian centroiding and fixed apertures. In Figure \ref{fig:phot_method_comp}, we show the performance of various fitting statistics for the four photometry methods versus the amount that the data were binned for one of the eclipses (data were binned over a range of timescales in our fitting procedure to best remove the intrapixel effect, as \citet{Deming2015} discuss). The standard deviation of normalized residuals, or SDNR, is a measurement of the scatter in the unbinned data after the fit has been calculated and applied, and the Gaussian-fixed method achieved a smaller value over all bin sizes compared to the other three methods. Variable aperture photometry is advantageous when there is significant variation in the shape of the apparent stellar image, such as can be produced by pointing jitter during an exposure.  However, when the exposure time is very short (as it is here), pointing jitter within an exposure is minimal.  In that case, variable aperture photometry can produce inferior results because the determination of the aperture radius is always affected by the photon noise in the image, resulting in spurious and unnecessary variations in the aperture radius.

All four photometry methods performed comparably well with the Allan variance slope, which is shown in the middle panel of Figure \ref{fig:phot_method_comp} and covered in more detail in Section \ref{subsec:apbin}. Lastly, the reduced-$\chi^2$ of fits (bottom panel of Figure \ref{fig:phot_method_comp}) made using the Gaussian-fixed photometry was better at all bin sizes than the other approaches. Similar results for other observations motivated us to restrict ourselves to only examining the photometry made using fixed apertures and Gaussian centroiding. 

To clean up the photometry, we discarded pixels that were discrepant by more than 6$\sigma$ of the standard deviation of the intensity of 9 neighboring pixels. If a given pixel was outside of this range, we replaced it with the median value of the 9-pixel neighborhood. We investigated the effect of our treatment of discrepant pixels in more depth. Strictly speaking, discrepant pixels should be zero-weighted in the photometry, not replaced by a median value \citep{Horne1986}. However, zero-weighting would be difficult to implement, given the structure of our photometry code, so we investigated the magnitude of possible bias introduced by median-filtering. We began with synthetic pixel data whose average value and scatter are the same as the real data. The frequency at which discrepant pixels occur is tracked by our photometry code, so we know on average how many pixels are replaced in the real data (see Table \ref{tab:data_table}). We chose that number of pixels at random, and calculate the effect of replacing them with a median filter in the synthetic data. Doing that 10,000 times, we calculated the average effect over the duration of the eclipse. We found that the effect is less than 0.5 ppm, and therefore of negligible significance to our analysis. We conclude that median-filtering of discrepant pixels is a valid method to clean the data prior to deriving eclipse depths.

We also discarded images that exhibited a significant amount of jitter in image position using a two-pass technique. The first pass made an absolute cut of images that exhibited image motion greater than 0.3 pixels, which eliminated large excursions. The second was a ``soft" cut that filtered images that showed jitter in disagreement with the standard deviation of the residuals of image position minus the median image position by more than 6$\sigma$. The percentage of discarded frames and corrected pixels are shown in Table \ref{tab:data_table}.

Because the current version of PLD is not designed to handle large amounts of image motion, we performed fits over a limited range of orbital phase around the eclipse. The size of this range is defined such that the amount of out-of-eclipse data is approximately equal to the amount of in-eclipse data. The first-order implementation of PLD used in this paper is valid for less than $\sim$0.2 pixels of image motion \citep[][]{Deming2015}, and limiting the phase range in this manner keeps image motion below this value for all eclipses.

\subsection{Fitting for Eclipse Depths}
\label{sec:Fitting for Eclipse Depths}
\subsubsection{Pixel-Level Decorrelation}
\label{subsec:PLD}
To remove systematic intra-pixel sensitivity fluctuations, we utilized the PLD framework presented by \citet{Deming2015}. Other methods currently used to correct for the intra-pixel effect depend on relating intensity fluctuations to the position of the stellar image on the detector (e.g. BLISS mapping [\citealp{Stevenson2012}], used by D16). As \citet{Deming2015} discuss, the position of the stellar image is a secondary data product, ultimately derived from the intensities of individual pixels. PLD utilizes the {\it primary} data, the intensities of individual pixels, to correct intra-pixel sensitivity variations.

PLD is capable of removing red noise that can frustrate other methods. \citet{Ingalls2016} compared seven different Spitzer decorrelation methods, and found that PLD (as well as BLISS mapping used by D16) was among the three methods that came closest to finding the true eclipse depth when applied to synthetic Spitzer data (however, it should be noted that six of the seven methods compared in \citet{Ingalls2016} produced results that were indistinguishable within the expected measurement error when applied to real and synthetic data). The PLD technique has been used successfullly in a number of recent \textit{Spitzer} analyses \citep{Kammer2015,Buhler2016,Fischer2016,Wong2016,Dittmann2017,Kilpatrick2017}. PLD is also the conceptual basis for the highly successful EVEREST code \citep{Luger2016} that achieves high precision in decorrelating data from the K2 mission.

Other known systematic effects that PLD must correct for include a temporal ``ramp" effect, in which the reported intensities of detector pixels change rapidly until they reach a threshold value, and temporal baseline effects, in which the intensity gradually increases or decreases over the course of observations. To account for the transient temporal ramp, we simply avoided data near the start of observations (see Section \ref{sec:photometry}). To remove the second effect, we fit visit-long functions to the photometry (linear, quadratic, or exponential) and subtracted them from the photometry. Through testing with the Bayesian Information Criterion (covered in more detail in Section \ref{subsec:modelbics}), we found that the best fits are achieved through use of visit-long linear baselines. For these observations, we found that not fitting a baseline at all typically results in failure to find a regression solution. As a result, we are confident that linear baselines are the best option for removing temporal baseline effects.

\begin{figure}
    \centering
    \includegraphics[scale=0.4]{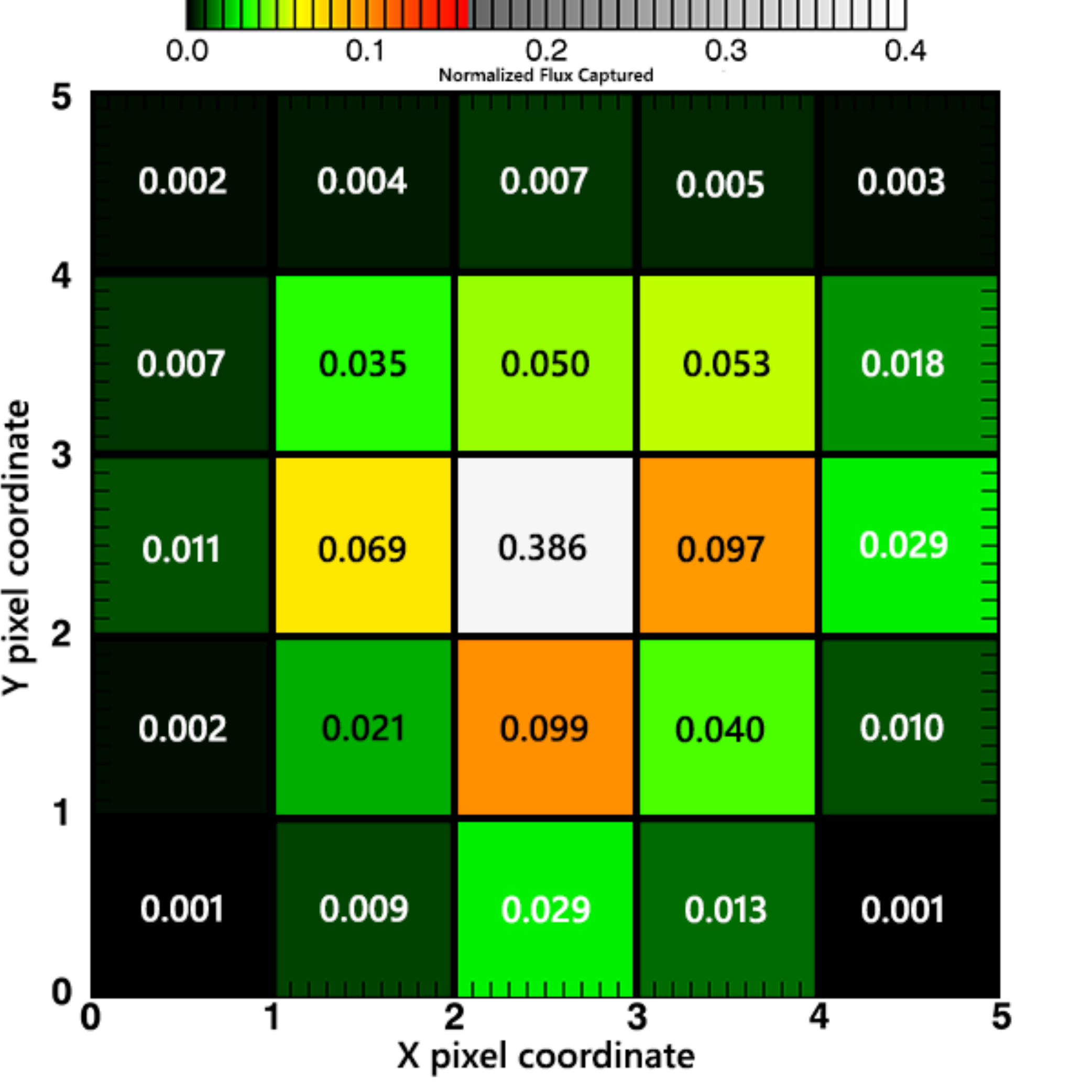}
    \caption{The fraction of normalized flux captured by each pixel in the 5x5 grid averaged over all observations. While the central 3x3 grid captures about the majority of the flux, we elect to use the entire grid to decorrelate the photometry because edge pixels capture about 15\% of the flux in the images, on average.}
    \label{fig:grid}
\end{figure}

Previous applications of the code used 9 \citep{Deming2015} or 12 \citep{Garhart2018} pixels, in accordance with the number of pixels that captured a significant portion of the stellar flux in those works. Here, we have updated PLD to work with anywhere from 1-25 pixels in a 5x5 grid surrounding the star. While in principle, the code can be can be used with any set of pixels in this grid, we expect the starlight to be distributed symmetrically about the central pixel upon which the star is placed in the image. This expectation, coupled with the fact that edge pixels in the 5x5 box captured ~15$\%$ of the total flux on average for these observations (see Figure \ref{fig:grid}), motivated us to use a 5x5 pixel grid to decorrelate the photometry.


\begin{figure}
    \centering
    \includegraphics[scale=0.5]{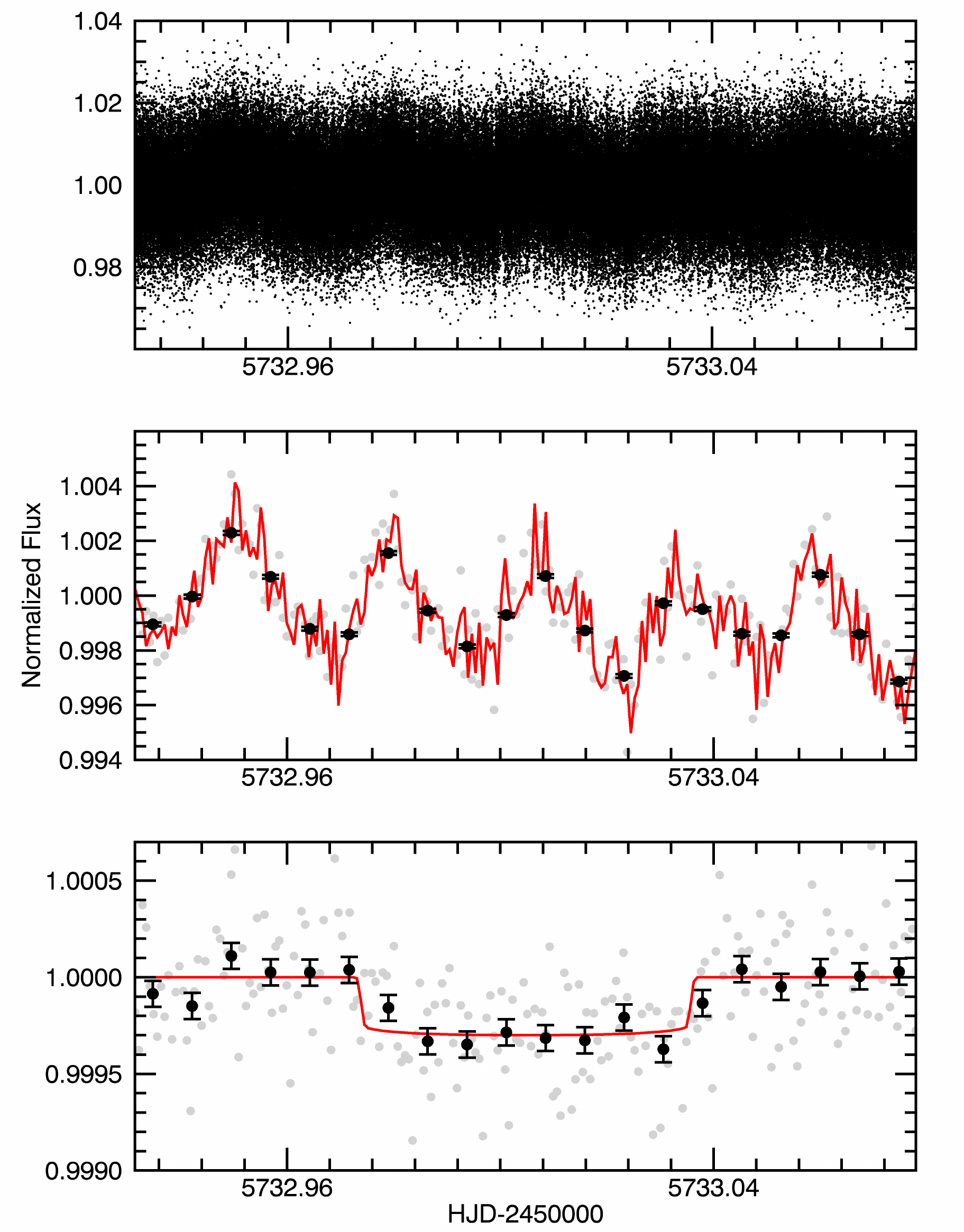}
    \caption{Application of PLD fit to photometry for the 2011 June 20 transit. \textit{Top:} Raw photometry, showing 239,443 unbinned points. \textit{Middle:} Binned photometry (grey and black circles) and overlaid best fit from PLD regression (red). The binning of the grey and black points are about 1 and 10.5 minutes, respectively, for clarity. This fit is reapplied to the unbinned photometry to examine the standard deviation of residuals on all time scales represented by the event. \textit{Bottom:} Binned photometry with intrapixel effect removed, with the best fit transit curve overlaid. Both grey and black points have the same binning as the middle panel.}
    \label{fig:TriplePlot}
\end{figure}

Figure \ref{fig:TriplePlot} shows a typical application of the regression fit on a 2011 transit, for the best binning and aperture size (see Sec. \ref{subsec:apbin}). In the figure, we show the process of going from raw photometry, to binned fit, to transit model with systematics removed. 

\subsubsection{Determining Central Phase}
\label{subsec:centralphase}
In the original version of PLD, the central phase of the eclipse was determined through fitting an eclipse model to unbinned data over a range of trial values, selecting the one that minimized $\chi^2$. We found that this method was unreliable for these observations of 55 Cnc \textit{e}, as the code could have difficulty in finding the small transit/eclipse events in the unbinned data. As a workaround, we set an initial guess for the central phase, based on literature values for the ephemeris \citep[][]{Endl2012}, and ran an instance of the MCMC on binned data to refine the central phase. This value was then used in subsequent instances of the code. 

For several of the eclipses, we found that even this workaround could not find a definite central phase, due to the shallow depths of the eclipses. We therefore elected to generate an ephemeris using the transit events, which are several times deeper than the expected eclipse depths. We used this ephemeris to calculate the expected eclipse times/central phases, assuming a zero-eccentricity orbit. Previous works \citep[][]{Nelson2014,Demory2016a} found eccentricities consistent with zero, despite the presence of four other planets in the system. We also calculated the tidal circularization timescale for 55 Cnc \textit{e}, using
\begin{equation}
    \tau_e = \frac{4}{63}Q(\frac{a^3}{GM})^{0.5}(\frac{m}{M})(\frac{a}{R_p})^5
\end{equation}

\citep{Goldreich1966}, where Q is the planet's specific dissipation function. Assuming a Q = 100 \citep{Gladman1996}, we find a tidal cirularization time of about 14,000 years, which justifies the assumption of a zero-eccentricity orbit.

\subsubsection{Selecting Aperture and Bin Sizes}
\label{subsec:apbin}
With the central phase provisionally determined, we varied the photometric aperture size, and binned the photometry over a range of time scales. Central to the PLD method is that it usually fits to data that are binned in time, as discussed in Sec.~3.1 of \citet{Deming2015}. \cite{Kammer2015} also discuss the rationale for binning in time. Briefly, if the scatter in the data is dominated by stochastic noise, it will be impossible to detect correlated noise and match it to our pixel basis vectors. By binning the data, we can diminish short-cadence white noise and expose long-cadence correlated noise. Fitting to binned data involves a loss of information (e.g., due to high frequency image jitter). However, we find that fitting to binned data produces a better match between the time scale of the fit and the dominant time scale of the intra-pixel fluctuations. Other authors \citep[e.g.][]{Deming2015,Buhler2016} have also used binned PLD fits very successfully.

As a function of both aperture radius and bin size, we searched for an optimum solution to the sum of the intra-pixel effect and the eclipse of the planet. We used 3 constant aperture radii (see Section \ref{subsubsec:nonevents}), and 112 bin sizes ranging from 2 to 2048 points-per-bin. For a given trial of bin size and aperture radius, we found the best fit by linear regression (solving Eq.~4 from \citealp{Deming2015}), holding the central phase of the eclipse fixed at the provisional value.  We then applied this trial fit to the \textit{unbinned} data, and we investigated the noise properties of those residuals as a function of time scale. We denote the standard deviation of the unbinned residuals as $\sigma(1)$. We binned the residuals over 2, 4, 8, 16, etc. points, until the effect of binning would reduce the number of residuals to $\le$ 16 points. We calculated the standard deviation of the residuals for each of these bin sizes, denoted as $\sigma(N)$.  

For a perfect fit that removes all but the photon noise, the standard deviation of the residuals (SDNR) should vary inversely with the square root of the bin size (a slope of -0.5 in $log_{10}$ space). We list the SDNR values of our fits compared to the photon noise in Table \ref{tab:data_table}. The relation between standard deviation and bin size is known as an Allan variance relation \citep[][]{Allan1966}. The original PLD method \cite{Deming2015} adopted the trial fit (i.e., bin size and aperture radius) that produced the minimum chi-squared in the Allan variance relation.  We have found it desirable to slightly modify that original criterion, as we now explain.

In choosing the best fit from different aperture radii and (especially) bin sizes, PLD effectively makes a compromise between mitigating the short-term scatter, represented by the standard deviations of unbinned normalized residuals (SDNR), and the longer-term red noise, represented by the standard deviations of the residuals binned to longer time scales. The standard deviation of the residuals itself has an uncertainty (i.e. the error on the error). That uncertainty increases with bin size because larger bins sample the error distribution with fewer points. Hence the chi-squared of the Allan variance relation is dominated by the smallest bins sizes. For that reason, we found that the original criterion used by \citet{Deming2015} (minimum chi-squared in the Allan variance relation) is virtually equivalent to choosing the solution with minimum unbinned scatter (i.e, minimizing SDNR). We therefore found it advantageous to modify the best fit criterion to choose the solution that has the minimum raw scatter in the Allan variance relation ($\sigma$AVR), after forcing the relation to pass through $\sigma(1)$. That is similar to the original criterion from \citet{Deming2015}, but with greater weighting for the larger bin sizes. Note that this fitting criterion is used (successfully) in Section \ref{subsubsec:nonevents} wherein we derived null-test eclipse depths to check our errors. 

In order to quantify the improvement of using our new version of the PLD code, we tested our new version versus the original code described by Deming et al. (2015). We used three representative eclipses (2012 Jan. 18, 2013 Jun. 15, and 2013 Jun. 29) and derived eclipse depths using the original code and our new code applied to the same photometry files, using the same degree of data binning. We also constrained the central phase of the eclipse to be the same in each code. We found that the new code reduces the single-point photometric scatter in the residuals (data minus fit) from 1.27 times the photon noise to 1.11 times the photon noise.  Given the large number of data points (239,000), that degree of reduction in the photometric scatter easily justifies (i.e, via a BIC analysis) the inclusion of the additional basis pixels in the PLD solutions. The average error on the eclipse depth, from the MCMC posterior distributions, is reduced from 50 ppm produced by the original code to 39 ppm from our new code. The average eclipse depth from the two codes was the same to within 1.2 times the error of the mean.

\subsubsection{Stability of Eclipse Depth Fits}
\label{subsubsec:stability}

In testing these observations, we found that many combinations of bin and aperture size produce comparably good $\sigma$AVR scores for a given eclipse. The eclipse models associated with those combinations, however, could have eclipse depths that differed by 10s of parts-per-million between each other (consistent with their errors), differences that could be significant when searching for potential variability between such small events. The distribution in eclipse depth for similar $\sigma$AVR scores is a result of the fact that $\sigma$AVR is not a rigorous statistical criterion. Instead, it's an attempt to get a ``broad-bandwidth" solution by trying to minimize both red and white noise through binning. Because of this, we selected our best-fit eclipse depth for each observation by examining a distribution of solutions, chosen by their $\sigma$AVR scores.

After running the regression for all bin and aperture sizes, we fit a Gaussian to the distribution of $\sigma$AVR values, which we refit after removing 3$\sigma$ outliers. We selected the center of this Gaussian as a cutoff value, and we then focus on solutions with $\sigma$AVR scores better than the cutoff. There are typically $\sim$100 solutions left after removing the bad solutions, and the \textit{eclipse depths} of these best solutions were approximately Gaussian-distributed for each eclipse. To determine the depth that we quote as our result, we fit a Gaussian to this distribution, selecting the combination of aperture and bin size that produced the eclipse depth that is closest to the central value. This approach allowed us to find a solution that is the most stable in the sense of representing the best fits found by the code, and prevented us from reporting an outlier as the best eclipse model. 


\subsubsection{MCMC}
\label{subsec:mcmc}
After finding the best combination of bin and aperture size, we held the binning and aperture constant, and refined the fit using a Markov Chain Monte Carlo procedure \citep[][]{Ford2005}. While we report the depth associated with the \textit{regression} solution, the MCMC allowed us to explore a high-dimensional parameter space (our models use $\sim$30 parameters when including the pixel coefficients) which accounted for possible correlations between fitting parameters and defines the error on the eclipse depth via a Gaussian fit to the posterior distribution.

The MCMC utilized the Metropolis-Hastings algorithm with Gibbs sampling.  At each step, we adjusted the eclipse depth, select orbital parameters (for transit fits), temporal baseline and pixel coefficients. Each chain consists of 500,000 steps, and we ran multiple chains so as to verify convergence through the Gelman-Rubin statistic, R \citep[][]{GelmanRubin1992}. 

Comparing our error estimates from the MCMC with those found by the best regression solution chosen in Section \ref{subsec:apbin}, we found that the MCMC errors are on average 13$\%$ higher than the regression errors for the eclipses.

We also tested initializing the MCMC with different solutions than the one chosen by our method of selecting a representative fit from the distribution of best $\sigma$AVR scores. These fits are typically within 2 ppm of the ``best" solution, but use different aperture and bin sizes. We find that the different initializing solutions did not produce significantly different errors from the solution chosen by our method. We therefore conclude that the errors are stable in the sense that they are not sensitive to the details of the fit.

\subsubsection{Independent Check on Errors}
\label{subsubsec:nonevents}
As stated above, our errors on the eclipse depths are determined using the MCMC simulations described in the previous section. However, deriving realistic errors is critical to the inference that the eclipse is variable, and a key feature of our analysis is an independent check on our derived errors.

To make this check, we identified eight regions in the 2013 eclipse data (two for each of the four eclipses) that were independent of each other and do not overlap with the actual eclipse. In these regions, we expect to derive eclipse depths that are consistent with zero within our error bars, because we know that they do not contain the actual eclipse. A critical test of the errors is that the derived ``eclipse" depths for these regions should scatter around zero with a dispersion that is consistent with their errors. 

We ``fit'' these non-eclipse regions following the methodology of Sections \ref{subsec:PLD}-\ref{subsec:mcmc}, locking the central phase of the model in each case. From this test, we found that if we restrict ourselves to using large photometric apertures (with radii of 3.1, 3.3, and 3.5 pixels, the same used in our eclipse analysis), we indeed find eclipse depths that vary around zero to a degree consistent with their derived errors (calculated through the standard deviation of in-eclipse and out-of-eclipse residuals, added in quadrature). The restriction to the larger aperture sizes makes sense given our choice of using all basis pixels in a 5x5 pixel grid surrounding the star: these apertures are well matched to the area of this basis pixel grid. Consequently, we limit ourselves to using these three apertures when fitting the actual transits and eclipses. 

\begin{figure}
    \centering
    \includegraphics[scale=0.5]{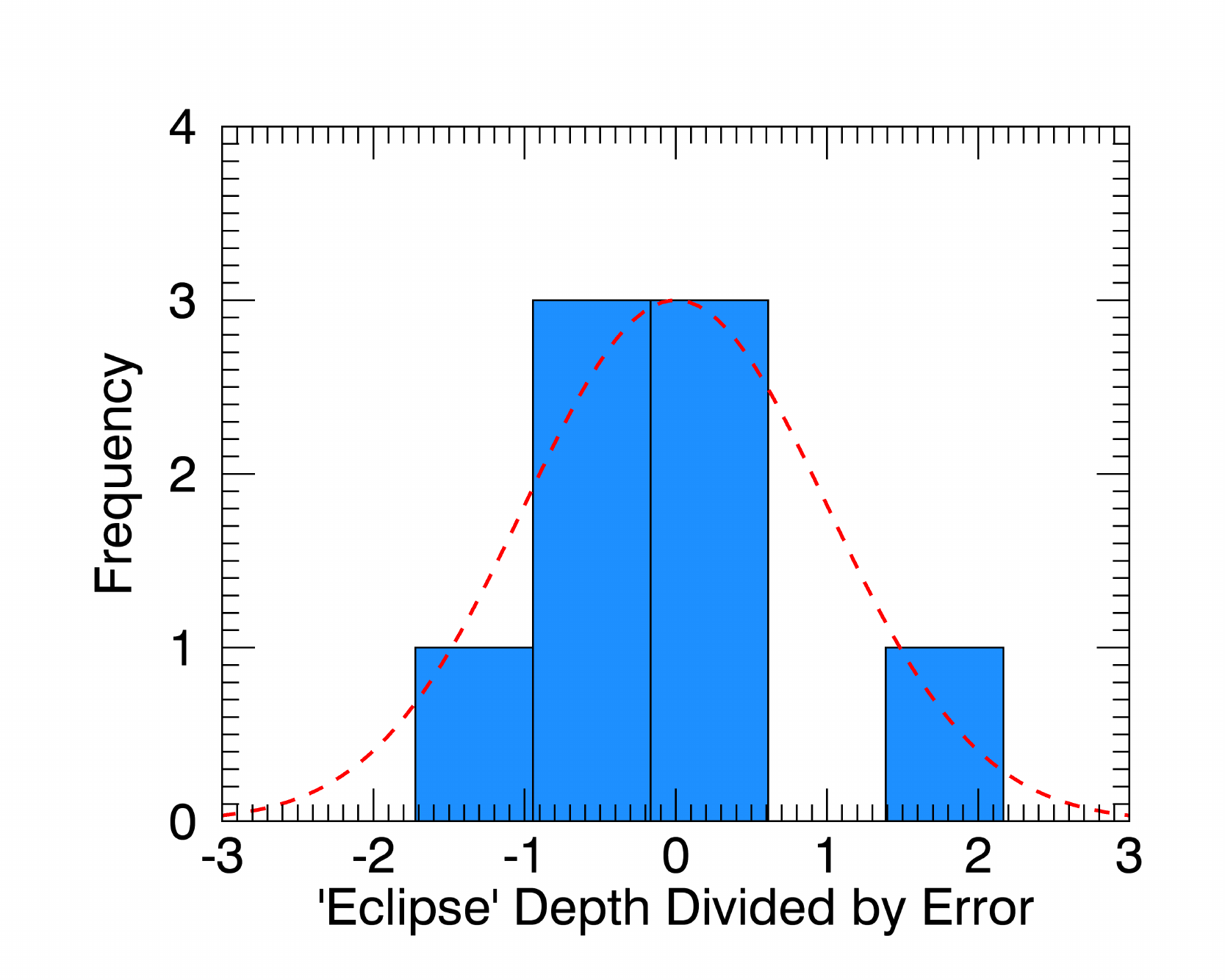}
    \caption{Results of testing the realism of our eclipse depth errors. The abscissa plots the distribution of each ``eclipse" depth divided by its error, with the histogram on the ordinate. If the errors are realistic, then the histogram should be consistent with a Gaussian distribution having a standard deviation of unity, overplotted on the histogram. In spite of our limited number of samples (8), this test shows that our errors are realistically determined.}
    \label{fig:noneclipse}
\end{figure}

The results of the fits to the non-event regions are shown in Figure \ref{fig:noneclipse}, which plots the distribution of eclipse depth divided by its derived error. If our errors are realistic, that distribution should be consistent with a Gaussian having standard deviation of unity. The distribution only contains eight samples, because we are limited by the amount of independent data regions that are available. Nevertheless, the distribution is approximately centered on zero with a dispersion consistent with unity. Our 1$\sigma$ error bars, which average to 34 ppm for these data, encompass all of the non-event results except two. The first of these is within 1.3$\sigma$ of zero, which we expect with a frequency of once every five measurements. The second is within 1.7$\sigma$ of zero, which is expected once every 11 measurements. 

Moreover, performing an Anderson-Darling test for normality \citep[][]{Stephens1974}, we find a p-value of 0.43, implying that our results for the non-event regions are consistent with a Gaussian having a standard deviation of unity. We conclude that our fitting procedure produces realistic errors, suitable for drawing inferences as to the variability of the eclipse.

\section{Results}
\label{sec:Results}

\subsection{Transit Analysis}
\label{sec:transit analysis}
 Early in our analysis, we found that the eclipse observations poorly constrain the planet's orbital parameters which can directly affect the retrieved eclipse model. As a result, we hold these parameters constant in our eclipse fits. However, we first needed the best possible estimates of these values, and because using different sets of published orbital parameters \citep[e.g.][]{Demory2016a,Demory2016b,Baluev2015,Nelson2014} gave slightly different values for inclination and central eclipse times, we elected to determine our own. The expected transit depths are significantly larger than the eclipse depths (by a factor of $\sim$4), and therefore constrain the planet's orbital parameters much more effectively. We began our analysis by examining the transit observations, to obtain self-consistent orbital parameters that would be used when fitting the eclipse data. 

\begin{figure}
    \centering
    \includegraphics[scale=0.5]{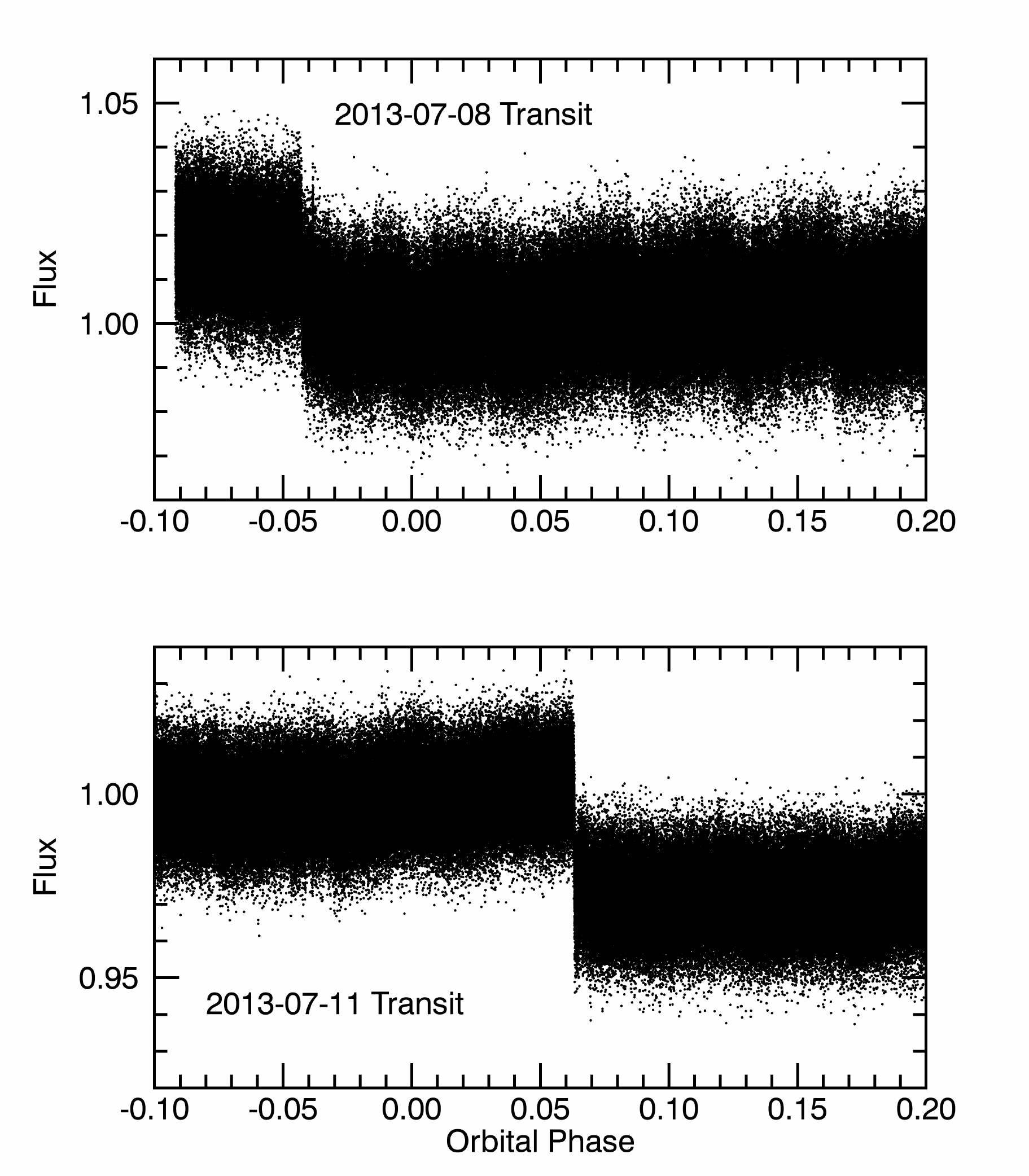}
    \caption{Raw photometry for the 2013-07-08 and 2013-07-11 transits. The sudden shifts in the photometry values are due to shifts in the stellar image on the detector, and limit PLD's ability to fit the data accurately.}
    \label{fig:photjump}
\end{figure}

We found that the photometry of both the 2013 July 8 and 2013 July 11 transits contain a sudden significant discontinuity mid-transit, which we show in Figure \ref{fig:photjump}. Further investigation showed that this was because the stellar image shifted suddenly on the detector in both cases. The 2013 July 8 transit shifted by about 0.3 pixels in the x-direction, and we find that the transit in this case is recoverable (with a larger error bar than our other results). The 2013 July 11 transit, however, shifted by almost 0.8 pixels in the x-direction, a jump that pulled a significant portion of the stellar PSF outside of our 5x5 pixel grid. Even expanding our grid with a sixth column, we found this observation to be intractable, with the regression and MCMC unable to find a similar transit depth. Rather than updating PLD to work for such large image motion, which is beyond the scope of this work, we chose to ignore this observation for the purposes of our analysis. 

 \begin{table}
        \centering
        \caption{Transit depths found in this work and D16. Our results share a correlation coefficient of 0.67 with D16's values.}
        \label{tab:transitdepths}
        \begin{tabular}{lcc}
            \hline
            Date & This Work (ppm) & Demory et al. 2016 (ppm)\\
            \hline
            2011 Jan 6 & 380 $\pm$ 36 & 484 $\pm$ 74\\
            2011 Jun 20 & 300 $\pm$ 35 & 287 $\pm$ 52\\
            2013 Jun 21 & 301 $\pm$ 38 & 325 $\pm$ 76\\
            2013 Jul 3 & 335 $\pm$ 36 & 365 $\pm$ 43\\
            2013 Jul 8 & 460 $\pm$ 77 & 406 $\pm$ 78\\
            \hline
            Weighted Average & 336 $\pm$ 18 & 360 $\pm$ 26\\
            \hline
        \end{tabular}
\end{table}
Due to computational limitations, we analyzed each of the five remaining transits separately. We allowed PLD to determine the best binning/aperture combination through our method outlined in Section \ref{subsec:apbin}. In the MCMC, we fit for the planet/star radius ratio $R_p/R_*$, orbital inclination $i$, quadratic limb darkening coefficients $u_1$ and $u_2$, the multiplicative coefficient of our linear ramp, and mid-transit time. We adopted the published value of $i$ from \citet{Demory2016b} as our starting value for inclination, and placed a Gaussian prior on this value based their computed error. We also priored the limb darkening coefficients $u_1$ and $u_2$ about their expected values from the tables of \citet{Claret2011} for the published stellar parameters \citep[][]{von Braun2011}. The convergence of these chains was determined by checking that the Gelman-Rubin statistic R \citep[][]{GelmanRubin1992} for the various parameters was less than 1.1. 

\begin{figure*}
    \label{fig:transmodels}
    \centering
    \includegraphics[scale=0.3]{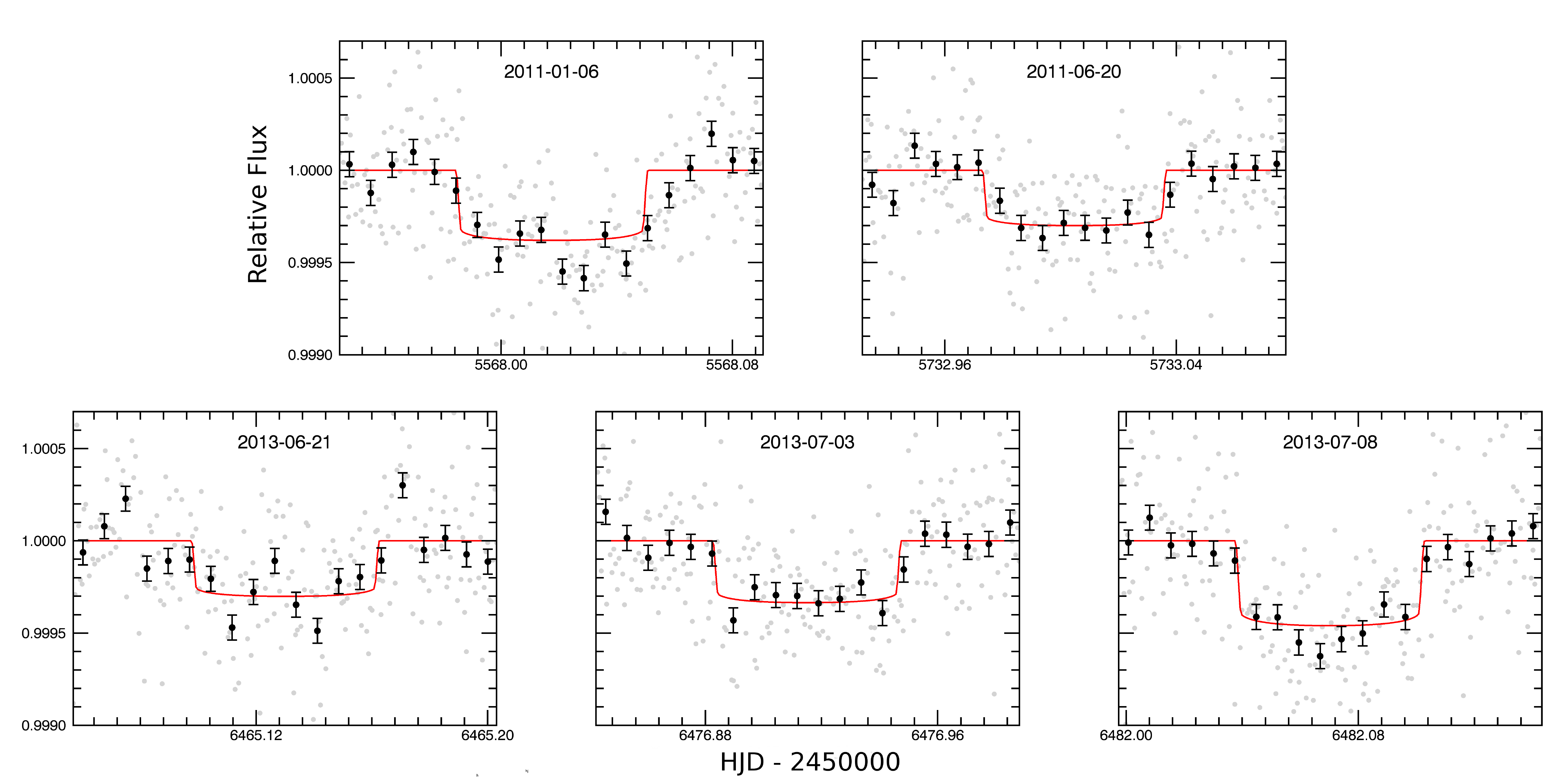}
    \caption{Best fit MCMC transit models. 2011 observations shown on top, 2013 observations shown on bottom.}
\end{figure*}

To determine the uncertainty on our regression transit depths, we performed a Gaussian fit to the combined MCMC posterior distribution of $R_p/R_*$ for each transit. The resulting depths/errors can be found in Table \ref{tab:transitdepths}, along with D16's transit results. Models of each transit can be found in Figure \ref{fig:transmodels}. We calculated the correlation coefficient between our transit depths and D16's and found a value of $\sim$0.67, indicative of positive correlation between the results. We also calculated the weighted average of our transit depths, and found that it agrees well with D16's weighted average.

We tested the sensitivity of the transit depth errors measured from the MCMC to the priors by doubling the prior widths on $i$, $u_1$, and $u_2$, and re-running three chains for each transit. We found that the errors measured from these runs differ by at most 2 ppm from those inferred from the original runs. We therefore conclude that the errors we report on transit depth are insensitive to the imposed priors.

We adopted a similar procedure for determining the planet's orbital parameters. We fit a Gaussian to the combined posterior distributions for central phase for each transit and use the value of each to perform a linear least squares regression to determine P and T0. For inclination, we combined the posterior distributions of all the transit observations and performed a Gaussian fit. The values we determined for these parameters can be found in Table \ref{tab:transparams}. We note that our values for P and i are within the 1$\sigma$ errors of the values published in \citet{Demory2016b}.

\begin{table}
        \centering
        \caption{Self-consistent parameter values as determined by performing MCMC procedures on transit observations. $R_p$ is calculated using the weighted average transit depth (see Table \ref{tab:transitdepths}), and assumes the published stellar radius from \citet{von Braun2011}.}
        \label{tab:transparams}
        \begin{tabular}{lc}
            \hline
            Parameter & Value\\
            \hline
            P (days) & 0.7365454 $\pm$ 0.0000003\\
            $T_0$ - 2,400,000 (HJD) & 55733.0037 $\pm$ 0.0002\\
            $R_p$ ($R_\oplus$) & 1.89 $\pm$ 0.05\\
            \hline
        \end{tabular}
\end{table}

Our value for T0, however, is a little over 1$\sigma$ away from the value reported in \citet{Demory2016b}. This is because our code finds that the first two transit events occur 16 and 12 minutes before the times expected from that paper's ephemeris, respectively. We tested initializing the MCMC with the value expected from their ephemeris, but the code still converged to the earlier event times. We also checked our agreement for these two observations with the ephemeris from \citet{Winn2011}, made using observations with MOST. Those values predict transits within 2.2 $\pm$ 1.5 minutes and 1.0 $\pm$ 2.0 minutes of our own measurements, respectively.

We elected to continue with our P and T0 values to remain self-consistent, but note that the difference in ephemerides could be a source of systematic difference between D16's eclipse results and our own. Our ephemeris predicts the 2012 and 2013 eclipses to occur $\sim$11 minutes and $\sim$ 4 minutes earlier than D16's, respectively. 

Calculating the weighted average of our transit depths, we found a depth of 336 ppm. We use this value to calculate the planet's average radius from 2011 to 2013, which we report as $1.89 \pm 0.05 R_\oplus$, a value in agreement with \citet{Demory2016a,Demory2016b}.

\subsection{Eclipse Analysis}
\label{sec:eclipse analysis}
\begin{figure*}
    \centering
    \includegraphics[scale=0.57]{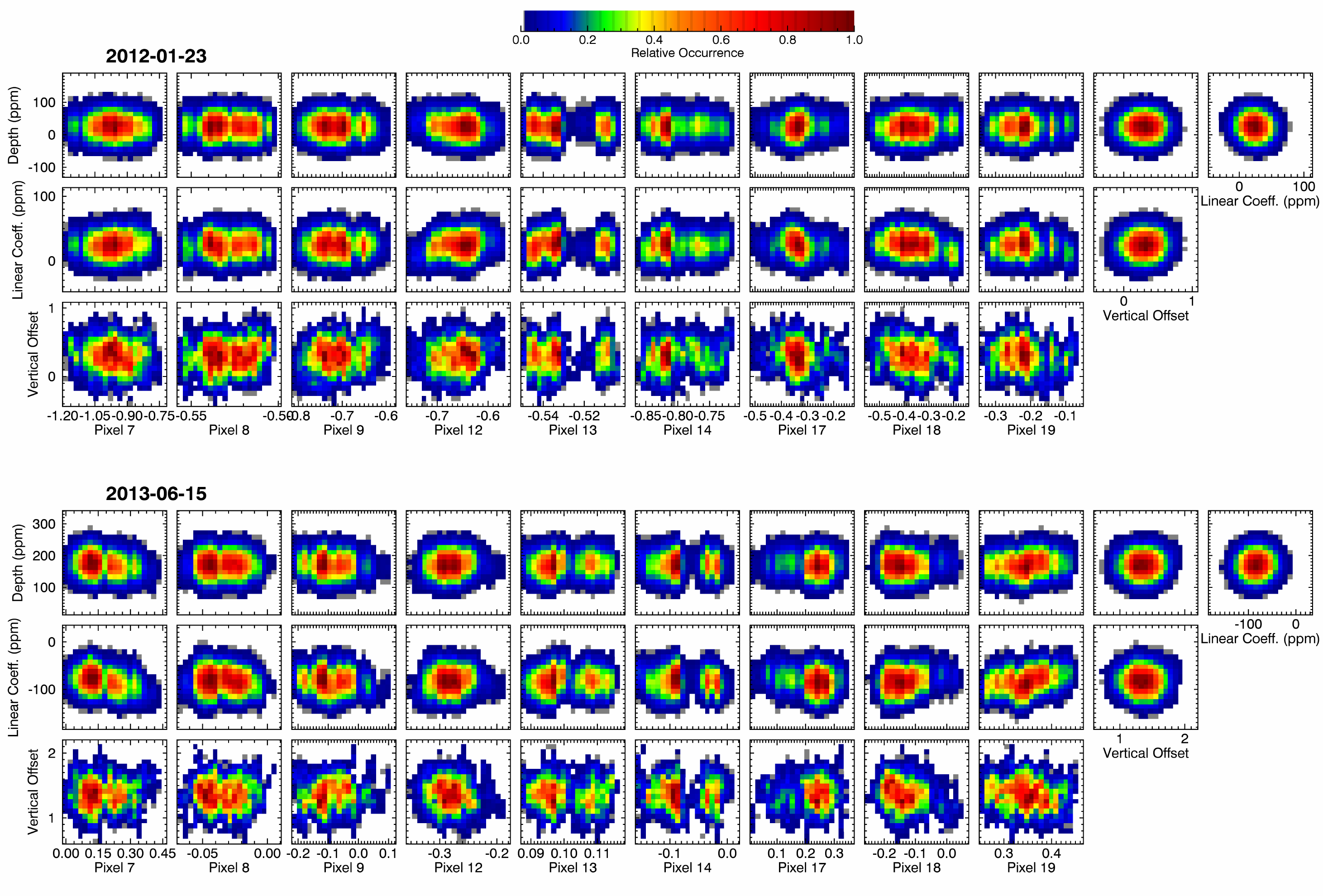}
    \caption{MCMC correlation plots for one 2012 and one 2013 eclipse. There is no correlation between any of the lightcurve parameters and any individual pixel position values.}
    \label{fig:nonpix}
\end{figure*}

For our eclipse analysis, we followed the same procedure as the transits to determine our regression solutions. In the MCMC, we allowed the pixel coefficients, linear slope coefficient, vertical offset, and eclipse depth to vary. Central phase, period, and inclination were locked to their expected values from our transit analysis (\ref{sec:transit analysis}), because these elements were not strongly constrained by the weak eclipses. $a/R_*$ was locked to the value reported by \citet{Demory2011}, and we assumed an eccentricity of zero for the planet's orbit. In Figure \ref{fig:nonpix}, we show corner plots of the non-pixel coefficients for one run of the MCMC for all eight eclipses. The lightcurve parameters show no correlation with individual pixel values, confirming that no degeneracy exists between the lightcurve solutions and the position of any specific pixel.

\begin{figure*}
    \centering
    \includegraphics[scale=0.4]{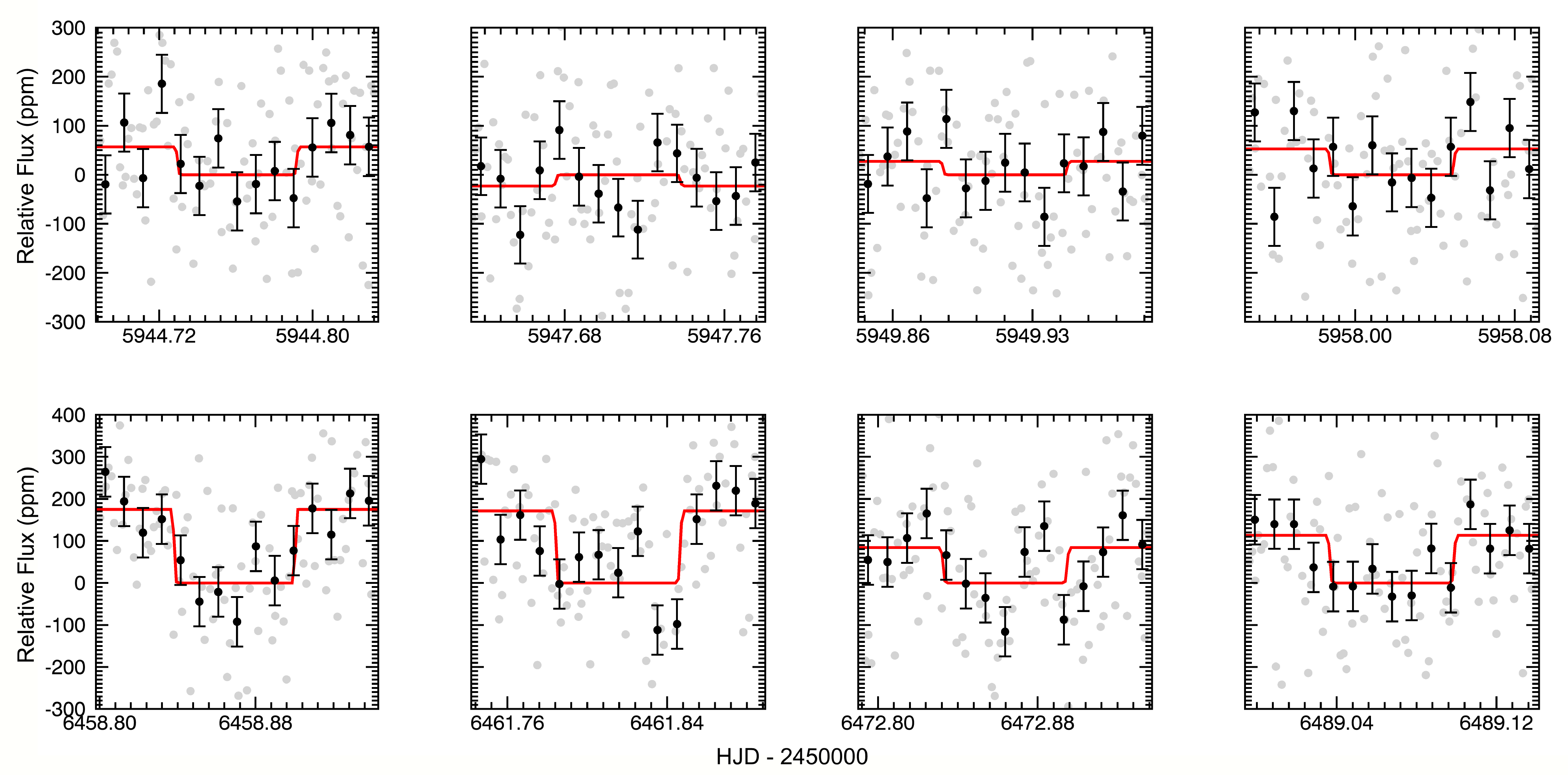}
    \caption{Eclipse light curve models. 2012 eclipses shown on top, 2013 eclipses on the bottom.}
    \label{fig:eclipsemodels}
\end{figure*}

The eclipse depths that we found are given in Table \ref{tab:eclipsedepths}, along with the reduced $\chi^2$ of the regression fits that initialized the MCMC. Lightcurve models for these fits are shown in Figure \ref{fig:eclipsemodels}.

\begin{table}
    \centering
    \caption{Eclipse depths found and reduced $\chi^2$ of regression fits found through PLD analysis. The depth given is our best regression result, while the error is derived from Gaussian fits to the posterior distribution of eclipse depths from the MCMC.}
    \label{tab:eclipsedepths}
    \begin{tabular}{|c|c|c|}
        \hline
        Date           & Depth (ppm) & $\chi^2_\nu$\\
        \hline 
        2012 Jan 18     & 57 $\pm$ 44 & 1.12 \\
        2012 Jan 21     & -23 $\pm$ 37 & 1.28\\
        2012 Jan 23     & 27 $\pm$ 35 & 1.24\\
        2012 Jan 31     & 53 $\pm$ 45 & 1.26\\
        2013 Jun 15     & 175 $\pm$ 38 & 1.13\\
        2013 Jun 18     & 171 $\pm$ 35 & 1.19\\ 
        2013 Jun 29     & 84 $\pm$ 39 & 1.20\\
        2013 Jul 15     & 113 $\pm$ 38 & 1.25\\
        \hline
    \end{tabular}
\end{table}

\section{Discussion}
\label{sec:Discussion}

\subsection{Transit Variability}
\label{subsec:transvar}
D16 found a 25\% variation in their transit depths in comparison to the mean value between 2011 and 2013 at the 1$\sigma$ level, which they considered to be a marginal detection of variability. With PLD's improvement over BLISS mapping's uncertainties for the transit observations (a 37$\%$ reduction, on average), we examined our own results to see if we find stronger support for transit depth variability. 

\begin{figure}
    \centering
    \includegraphics[scale=0.5]{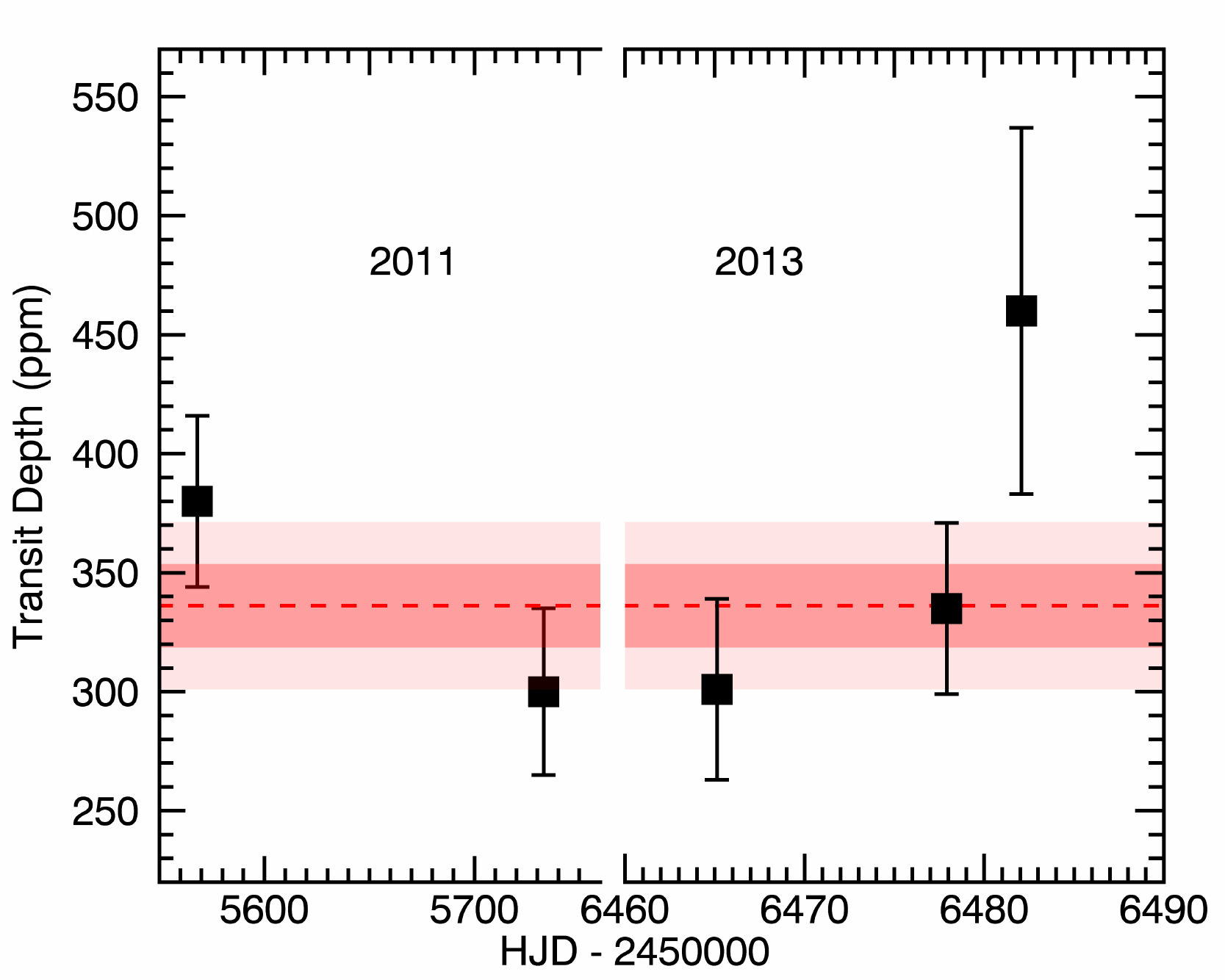}
    \caption{Our transit depth results versus time, with 2011 shown on the left and 2013 on the right. The 2$\sigma$ range of a flat line fit to the depths is shown in red.}
    \label{fig:transflat}

\end{figure}

We began by calculating the correlation coefficient of D16's transit depths and our own, finding a value of 0.67, indicative of agreement between the two sets of results. We then calculated a weighted average of our transit depths, giving $336 \pm 18$ ppm. This calculation represents the null hypothesis, that the transits can be represented by a single constant depth. We show our transit depths versus time with the 2$\sigma$ range of this flat line fit overplotted in Figure \ref{fig:transflat}. The non-variable model gives a $\chi^2$ of $6.0$ for our results, which, for $n = 4$ degrees of freedom, gives a p-value of about $0.2$. Our results for the first five transits, therefore, are consistent with a constant depth, as we would expect a higher $\chi^2$ value around 20$\%$ of the time due to random chance alone. On this basis, we conclude that our results do not provide evidence for different transit depths for 55 Cnc \textit{e} in 2011 compared with 2013. Therefore, both the results of D16 and this work find insufficient evidence to support a conclusion of year-to-year variability in 55 Cnc \textit{e}'s \textit{transit} depth.

However, the separate linear trends in the transit depth within each year found by D16 are still present in our results, though the variation falls within our 3$\sigma$ uncertainty envelope. Due to the short baseline of the observations for each year, we cannot rule out intra-annual variability, and future higher-cadence and/or longer-baseline transit measurements over $>$5 orbital periods would help to resolve whether these trends are robust or solely due to observational uncertainty.

\subsection{Eclipse Variability}
\label{subsec:eclipsevar}
One way to determine the reality of statistical fluctuations in 55 Cnc \textit{e}'s eclipse depth in the presence of possible analysis errors is to determine whether both D16's application of BLISS mapping and our own use of PLD produce correlated results when comparing eclipse-to-eclipse. We calculated the correlation coefficient between the sets of eclipse depths found through these two methods, and find a value of 0.90, which tells us that the two sets of results are significantly correlated. The high correlation between the results from PLD and BLISS mapping proves that the variation in eclipse depth is present \textit{in the data}, not in the analysis approach, and it reflects the agreement found between the two methods in \citet{Ingalls2016}.
    
Moreover, fitting our eclipse depths with a flat line gave a $\chi^2$ p-value of $8.3e-11$, indicating that a constant eclipse depth is an extremely poor fit to the results. We therefore conclude that the variability of 55 Cnc \textit{e}'s eclipse depth as reported in D16 is genuine. 

D16 reported an increase from an average 47 $\pm$ 21 ppm depth in 2012 to a 176 $\pm$ 28 ppm depth in 2013 at the 4$\sigma$ level. Having agreed that the eclipse depth is indeed variable, we turned our attention to modeling the depth versus time, to see if our results support the same average-increase model as proposed by D16.

\subsubsection{Heuristic Models}
\label{subsec:modelcalcs}

\begin{figure*}
    \centering
    \includegraphics[scale=0.5]{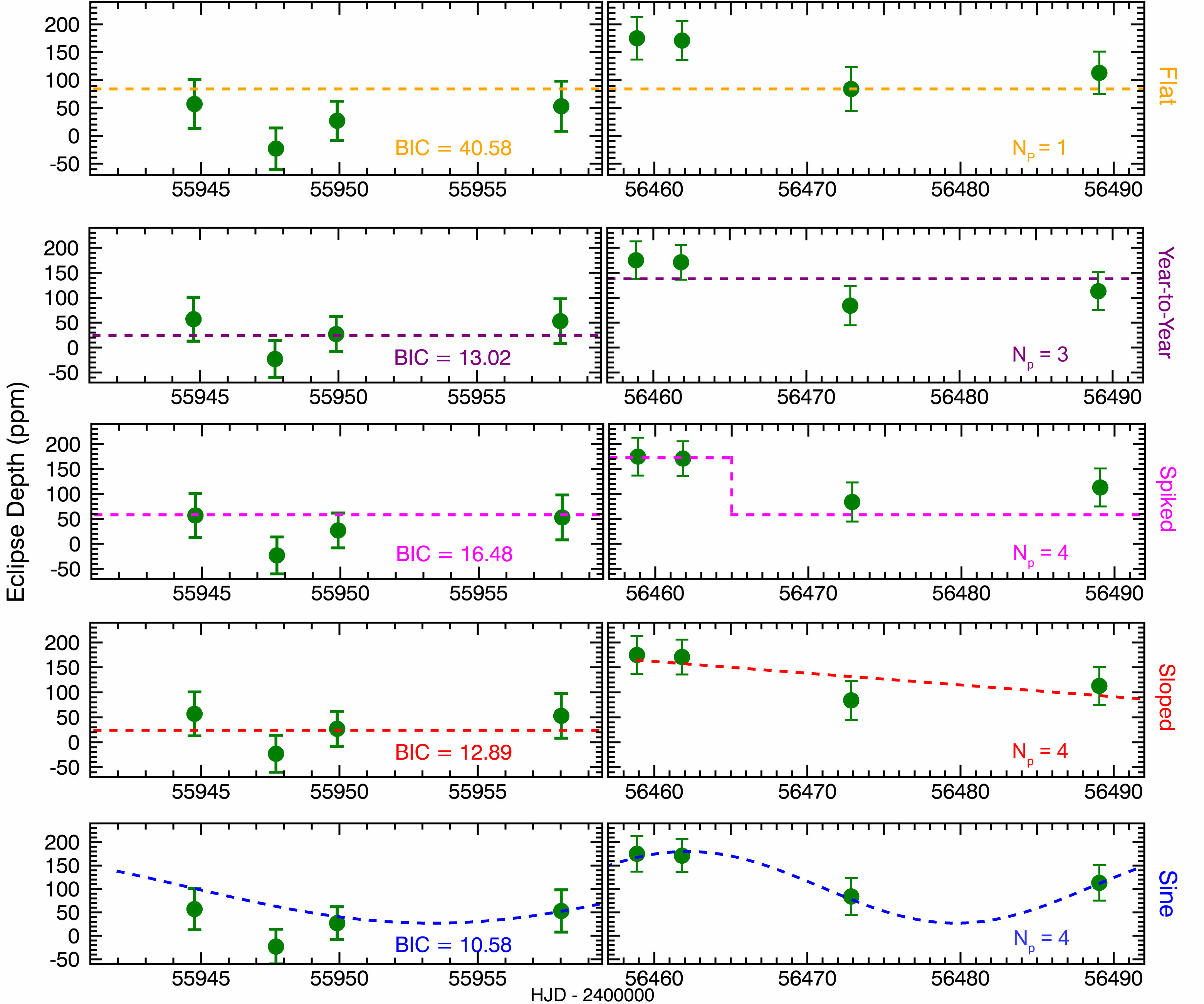}
    \caption{Eclipse depths versus time with our five different models overplotted. By calculating the Bayesian Information Criterion for each model (shown in the left panel of each model), we are able to conclude that a constant eclipse depth model fits the results the least well out of all five. The number of free parameters, $N_p$, is shown in the right panel of each model.}
    \label{fig:models}

\end{figure*}

We generated five different models to explain the pattern of eclipse depths versus time using Levenberg-Marquardt fitting in IDL, which we show in Figure \ref{fig:models}. We now describe each of these models in detail.

The first model is a flat eclipse depth, i.e. one that is not varying in time. This is the model that we would anticipate for a ``normal'' planet, as we would expect to see a mostly constant thermal flux from the planet versus time. There is only one free parameter for the flat model, that being the weighted average eclipse depth. We find a depth of $84 \pm 14$ $ppm$ best fits all the eclipses in the flat model. 

The second model is similar to the one proposed by D16, with year-to-year variability between the 2012 and 2013 eclipses. We fit constants to our depths from 2012 and 2013 separately to determine this model. The year-to-year model is characterized by three free parameters: an eclipse depth for both 2012 and 2013, and a time of transition. For 2012, we find a depth of $24 \pm 20$ $ppm$, and a depth of $138 \pm 18$ $ppm$ in 2013. 

The third model, which we refer to as the ``spike" model, uses a flat depth everywhere except for the first two observations in 2013, which we found were the two deepest eclipses. The four free parameters of the spike model are the baseline eclipse depth, the depth of the ``spike", and two transition times. For this model, we find a baseline depth of $58 \pm 16$ $ppm$, and a depth of $173 \pm 26$ $ppm$ for the first two observations in 2013. 

The fourth model, which we call the ``sloped" model, is similar to the third, but with a linearly decreasing slope in 2013. It also has four free parameters: the average 2012 eclipse depth, a time of transition, and a peak and slope value for the 2013 data. It uses the same 2012 value as the year-to-year model, and perform a linear fit to the 2013 data of the form $y = mx + b$. We find $m = -2.36 \pm 1.59$ $\frac{ppm}{day}$, and $b = 164.7 \pm 26.0$ $ppm$. 

\begin{figure}
    \centering
    \includegraphics[scale=0.4]{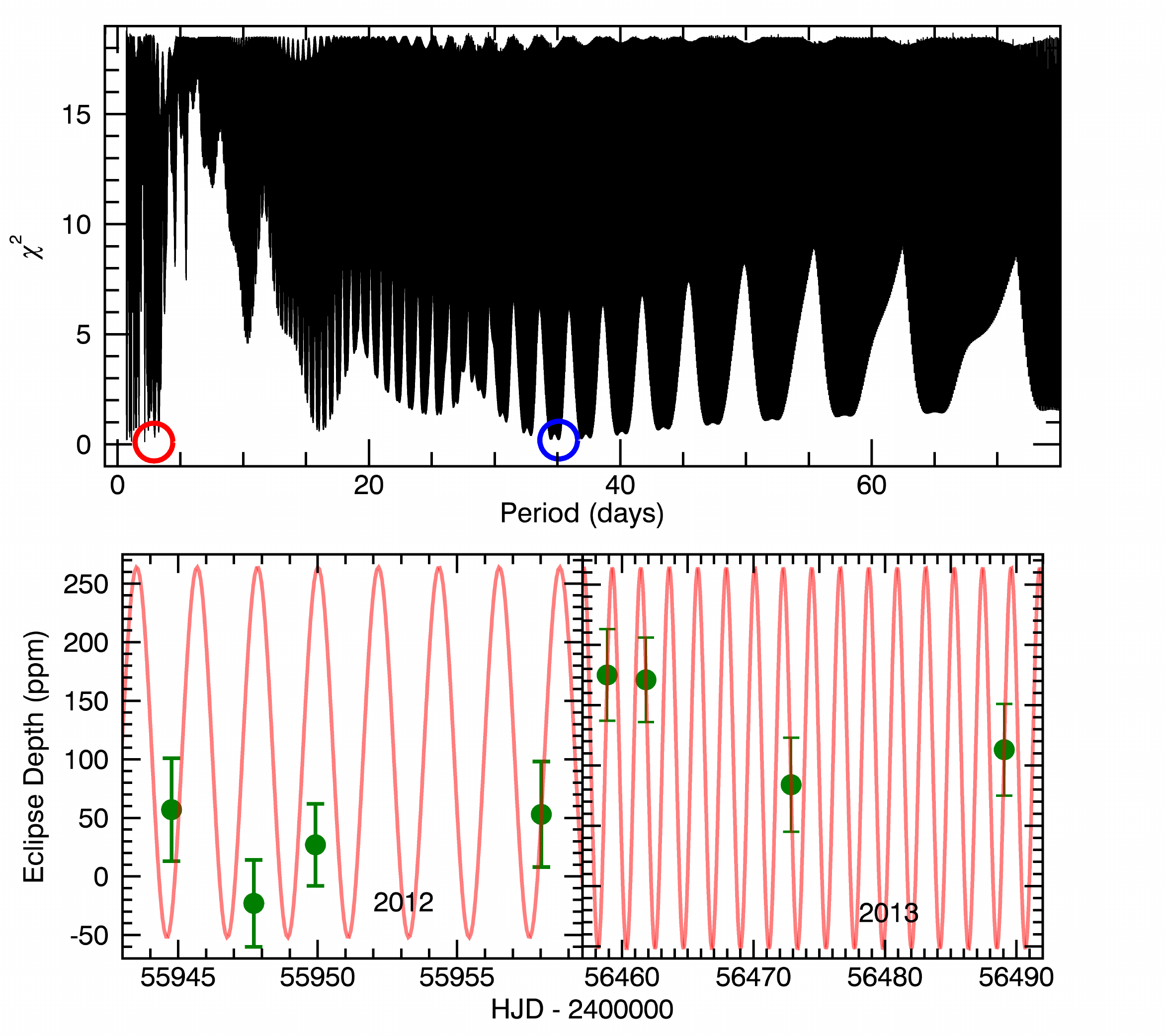}
    \caption{\textit{Top}: $\chi^2$ versus period for our sine wave model fits to the eclipse depths. A number of different periods give models with similar $\chi^2$, a result of the data being heavily aliased. \textit{Bottom}: Visualization of the ``best" sine wave fit (circled in red in the top panel), which clearly represents an unphysical model and heavily overfits the data. For the purposes of examining a sine model further, we use the best model with a period greater than 20 days. This fit is circled in blue in the top panel, and shown in the fifth panel of Figure \ref{fig:models}.}
    \label{fig:sineperiod}
\end{figure}

The final model is a simple sine wave, given by $Asin(2\pi ft + \delta) + c$, where $A$ is the amplitude of the wave, $f$ is the frequency, $\delta$ is a phase shift and $c$ is a constant offset from 0 (i.e. four free parameters). To determine this model, we fit a sine function to our best-fit eclipse depth results using Levenberg-Marquardt fitting, stepping over periods that ranged from half a day to about three years. At each frequency step, we fit for sine wave amplitude, phase offset, and constant offset, and recorded the $\chi^2$ of each fit. We plot the results in Figure \ref{fig:sineperiod}, with frequencies converted to periods in days. This model represents a periodic cycling of the planet's eclipse depth over time. Physically, a sine wave could be a useful approximation of any process (e.g. volcanism) that has a characteristic timescale, i.e. it represents a bandwidth-filtered stochastic process. 

Figure \ref{fig:sineperiod} shows that a number of different sine wave periods fit the data similarly well. This is because the data are heavily aliased, as our sampling rate does not permit us to distinguish between sine waves of many different frequencies. In the bottom panel of the figure, we show the sine model generated using the period circled in red in the top panel. Fluctuations of 300 ppm over a 2 day period seem unphysical, yet fitting a sine wave to our aliased data finds that this is the ``best" model. 

While we again acknowledge that the data are heavily aliased, we continue forward with a sine wave model because future data could better constrain the periodicity (if periodic fluctuation is actually the physical reality of the system). We pursue a sine model with a less-extreme period, electing to use the best fit with periodicity greater than 20 days. This fit is circled in blue in \ref{fig:sineperiod}. We found the model's parameters to be $A = 76.46 \pm 18.99$ $ppm$, $f = 0.179057$ $days^{-1}$, $\delta = 1.28 \pm 0.30$, and $c = 103.54 \pm 14.16$ $ppm$.

\subsubsection{Model BICs}
\label{subsec:modelbics}
As in other works \citep[e.g.][]{Gibson2010,Haynes2015}, we calculated the Bayesian Information Criterion \citep[][]{Schwarz1978} to assess the relative quality of fit between these models. The BIC is calculated through the formula $BIC = \chi^2 + k \cdot ln(n)$, where $k$ is the number of model parameters (1 for the flat model, 4 for the spike model, 3 for the year-to-year model, 4 for the sloped model, and 4 for the sine model) and $n$ is the number of data points (8 in our case). A lower BIC is preferred, with an improvement of 6-10 points representing strong evidence for one model over another, and more than 10 points representing very strong evidence \citep[see][]{Kass1995}. We list the BICs for each model in Figure \ref{fig:models}. 

Based on our analysis, we found that the flat model is clearly ruled out by its BIC score, despite using the fewest number of parameters. The remaining models all have BICs that are within 6 of each other, so we are unable to confidently distinguish between them. Because the year-to-year model has the fewest free parameters out of the variable models, we conclude that it is the best model we currently have for interpreting the planet's eclipse depth variability. While the sine model has the \textit{best} BIC score, we again emphasize that the fit was performed to heavily aliased data, and more observations would be required to determine if a sine model is realistic.

\subsubsection{Brightness Temperature Calculation}
\label{subsec:brighttemp}
Following the calculations of \citet{Crossfield2012}, we converted the maxima and minima of our five models into planetary brightness temperatures ($T_B$) using an observed spectrum of 55 Cnc. That paper demonstrated that the observed spectrum gives a significantly lower brightness temperature for 55 Cnc \textit{e} than would be calculated when treating the star as a blackbody. Using their measured flux density of $3.444 \pm 0.006$ Jy, we numerically evaluated the following integral for values of $T_B$: 
\begin{equation}
    \int B_\lambda(T_B)S_x(\lambda)d\lambda = \delta_{occ}\frac{R_*}{R_p}\int F_\lambda^*S_x(\lambda)d\lambda 
\end{equation}

...where $B_\lambda(T_B)$ is the wavelength version of Planck's law as a function of the planet's brightness temperature, $S_x(\lambda)$ is the relative system response in Spitzer IRAC Channel 2\footnote[1]{http://irsa.ipac.caltech.edu/data/SPITZER/docs/irac/
calibrationfiles/spectralresponse/}, $F_\lambda^*$ is the stellar flux density (converted to $W m^{-3} sr^{-1}$), and $\delta_{occ}$ is the planet's measured eclipse depth. 
The maximum and minimum values of $T_B$ for each model can be found in Table \ref{tab:brightnesstemps}. 

To examine the plausibility of our different models, we are motivated to explore what physical mechanisms could account for the brightness temperature variations suggested by the year-to-year and sine models.

\begin{table}
    \centering
    \begin{tabular}{|c|c|c|}
        \hline
        Model & Minimum $T_B$ (K) & Maximum $T_B$ (K)\\[2ex]
        \hline
    Flat & $1820^{+142}_{-147}$ & $1820^{+142}_{-147}$\\[1ex]
    Spike & $1550^{+175}_{-188}$ & $2642^{+240}_{-242}$\\[1ex]
    Year-to-Year & $1115^{+266}_{-415}$ & $2331^{+183}_{-185}$ \\[1ex]
    Sloped & $1115^{+266}_{-415} $  & $2571^{+242}_{-245}$\\[1ex]
    Sine & $1163^{+330}_{-628}$ & $2705^{+249}_{-252}$\\[1ex]
    \hline
    \end{tabular}
    \caption{Maximum and minimum brightness temperatures for each of five models. Temperatures were calculated using an observed stellar spectrum of 55 Cnc from \citet{Crossfield2012}.}
    \label{tab:brightnesstemps}
\end{table}

\subparagraph{Variability in Stellar Flux}
\label{subsubsec:starspots}
The most likely mechanism for variations in stellar flux on day-to-week timescales is the influence of star spots.  Analyzing 11 years of photometric observations of 55 Cnc, \citet{Fischer2008} found measurements covering two consecutive rotational periods ($\sim$80 days) that showed evidence of low-amplitude, short term-variability. This variability was attributed to star spots covering less than 1$\%$ of the star's surface, resulting in a brightness amplitude variation of 0.006 mag. While this variation could produce a signal of $\sim$200 ppm in transit/eclipse observations \citep[][]{Demory2016b}, the fact that \citet{Fischer2008} found 55 Cnc to be a quiet star makes this an unlikely scenario. Furthermore, the star would have to be variable on the time scale of the eclipse itself in order to affect our results. On this basis, we conclude that the star is not responsible for the observed eclipse depth variability. 

\subparagraph{Asynchronous Rotation with a Hot Spot}
\label{subpar:asynchronous}
Because the measured variability is not well accounted for by changes in the stellar flux, the variation must be due to changes in the planet's flux at 4.5 $\mu$m. If the variability is truly periodic, one of the easiest physical explanations to visualize is that 55 Cnc \textit{e} is rotating asynchronously while maintaining a hotspot (e.g., from volcanism) on one side of the planet. In this scenario, we would observe the occultation of the planet's hot, bright side in some observations, and its cooler, darker side in others. This would lead to more significant drops in relative flux during some observations than in others in a periodic fashion that matches the planet's spin. While we cannot decisively constrain the planet's rotation rate from our results, Figure \ref{fig:sineperiod} suggests periods between 15 and 80 days. 

A valid objection to this scenario would be the expected timescale for tidal locking for 55 Cnc \textit{e}, which would cause us to see the same face of the planet before/after eclipse in every observation. We can estimate this timescale from Eq. 9 given in \citet{Gladman1996}:
\begin{equation}
    \tau_{lock}=\frac{\omega a^6 I Q}{3Gm_{s}^2k_2R_p^5}
\end{equation}
where $\omega$ is the spin rate, a is the radial distance from planet to star, I is the planet's moment of inertia about the spin axis, Q is the planet's specific dissipation function, G is the gravitational constant, $m_S$ is the mass of the star, $k_2$ is the tidal Love number of the planet, and $R_p$ is the mean radius of the planet. We use the estimate of Q = 100 from \citet{Gladman1996} for satellites in our solar system, and calculate $k_2$ by

\begin{equation}
    k_2 = \frac{3/2}{1+\frac{19\mu}{2 \rho g R_p}}
\end{equation}

\citep[][]{Burns1977} where $\mu$ is estimated as $3$x$10^{10} \frac{N}{m}$ for rocky bodies \citep[][]{Gladman1996}, $\rho$ is the average planetary density (using $M_p$ from \citet{Demory2016b} and $R_p$ from this work), $g$ is the surface gravity of the planet, and $R_p$ is the planetary radius. Due to the small semi-major axis and high bulk rigidity of 55 Cnc \textit{e}, de-spin times are on the order of hundreds of years, which makes unsynchronized rotation an unlikely explanation for variability.

Alternatively, it is possible that the planet could become captured in a higher-order spin-orbit resonance as it spun down. Mercury, with its 3:2 resonance \citep[][]{Pettengill1965}, is an example of such a configuration in our own solar system. Further observations of 55 Cnc \textit{e} will be required to resolve this possibility.

\subparagraph{Equilibrating Refractory Particulates}
\label{subsec:volcanism}

If we do not expect significant flux changes from the host star and we assume that the planet is tidally locked, then thermal variability must be occurring on the face of 55 Cnc \textit{e} that we observe before/after eclipse. Due to its close-in orbit and proximity to its host star, the planet most likely has a molten surface and may experience intense tidal forces, both of which could drive volatile loss through either surface evaporation and/or volcanic activity \citep[][]{Schaefer2009}. Volcanic eruptions or a volatile atmosphere could result in particulates either lofted or condensed high above the surface \citep[][]{Mahapatra2017}. D16 examined this as a source of thermal variability, with volcanic plumes raising regions of the planet's photosphere to higher, cooler regions in 2012, which lead to smaller eclipse depths (i.e. less thermal emission) that year. 

Here, we explore a similar but more general concept. For each of our models (except the flat model, for which this doesn't apply because the depth is constant), we assume that our measured value of $T_{b,max}$ represents the temperature of the radiating planetary surface, and we check to see if the planet's equilibrium temperature (due to stellar insolation alone) is consistent with this measurement. If the planet has a molten surface or is volcanically active, it may outgas material that cools at a height above the planet, blocking out the hot surface and leading to lower temperatures. For each model, we check to see if $T_{b,min}$ can be achieved by this obscuring volcanic ejecta. 

To start, we determined the equilibrium temperature of 55 Cnc \textit{e} due to stellar insolation alone over a range of planetary albedos. We did this both for an efficient cooling case, in which the planet is able to radiate heat away over its whole surface, and an inefficient case, in which it can only radiate over half the surface. 

\begin{figure*}
\centering
\begin{tabular}{cc}
  \includegraphics[width=65mm]{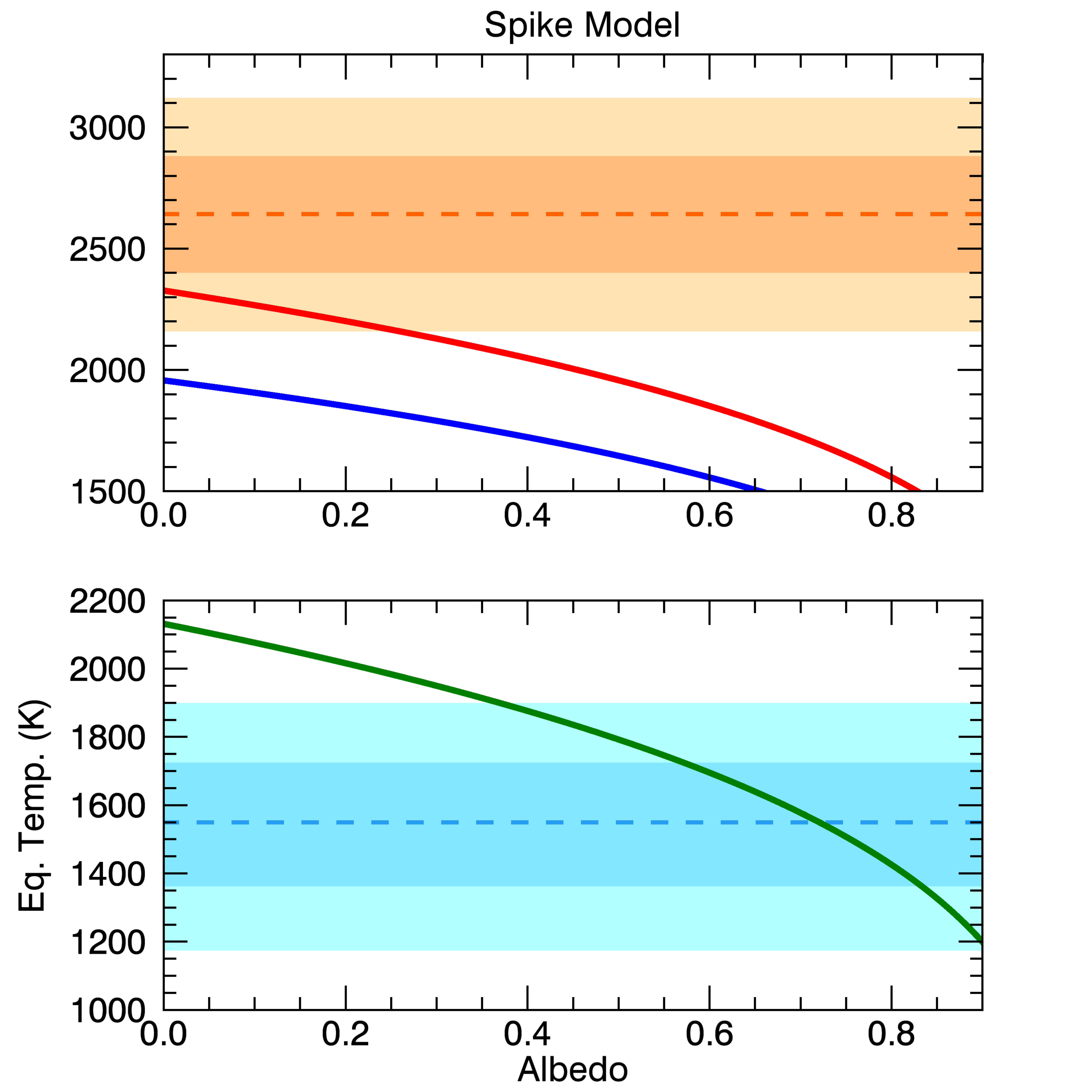} &   \includegraphics[width=65mm]{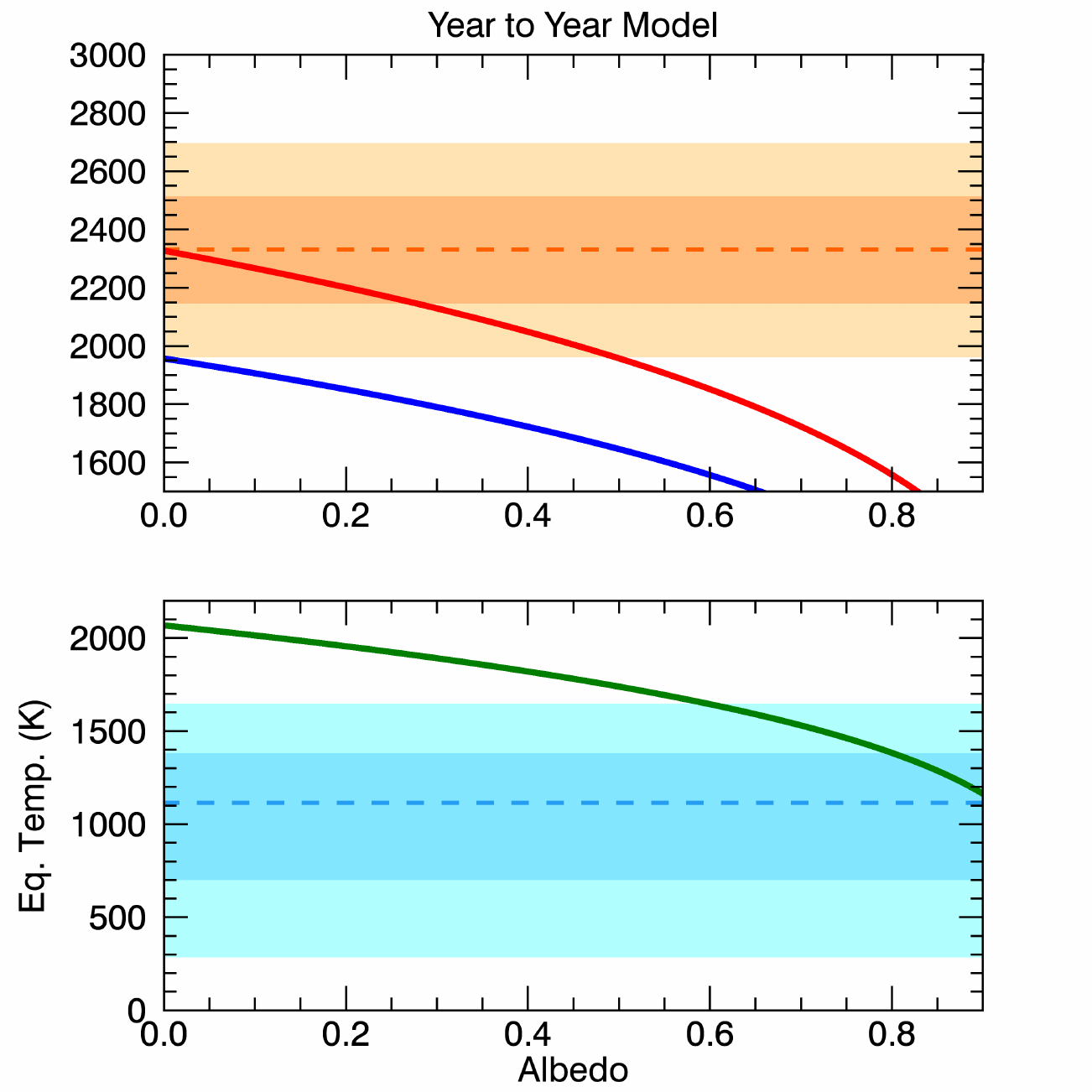} \\ 
  \\
 \includegraphics[width=65mm]{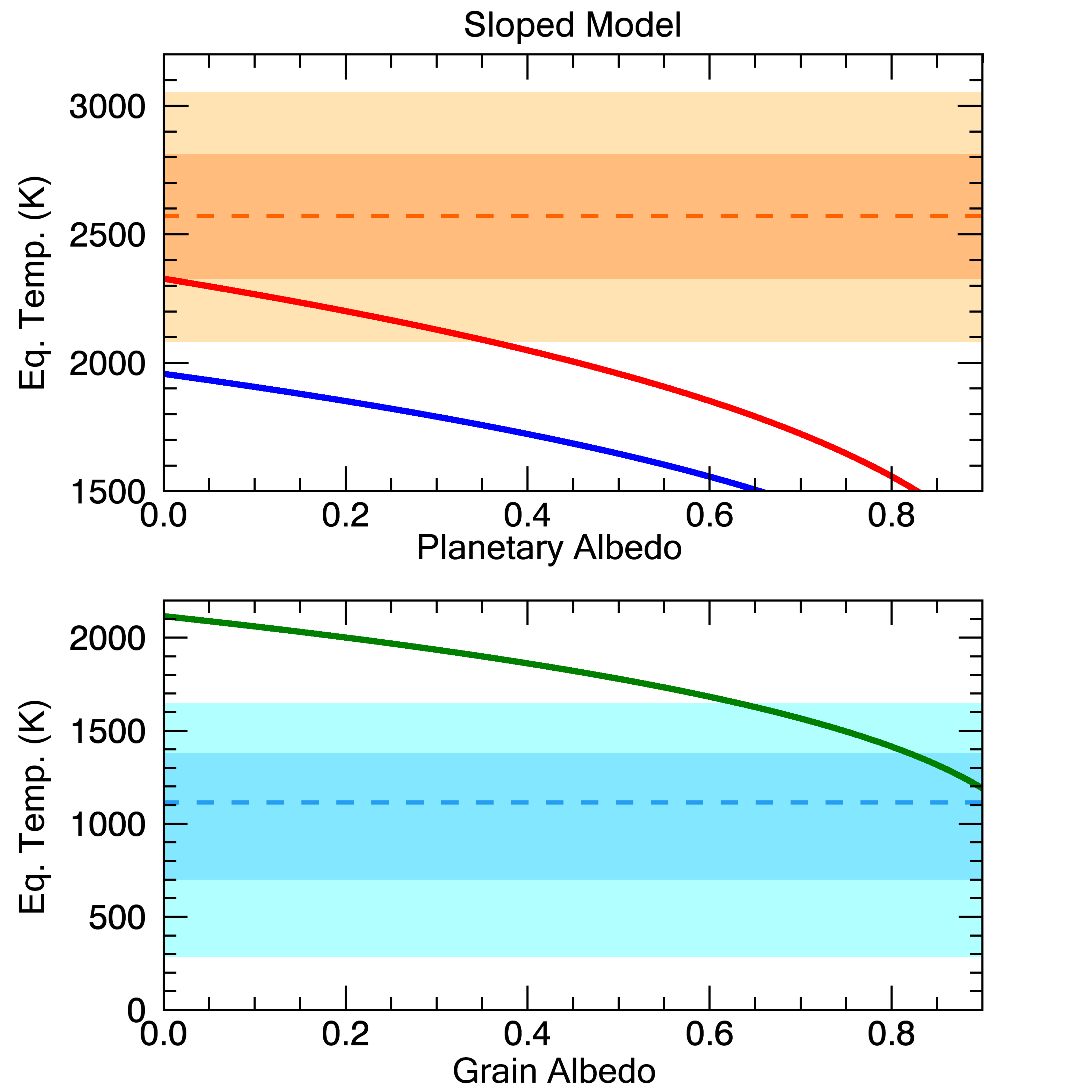} &   \includegraphics[width=65mm]{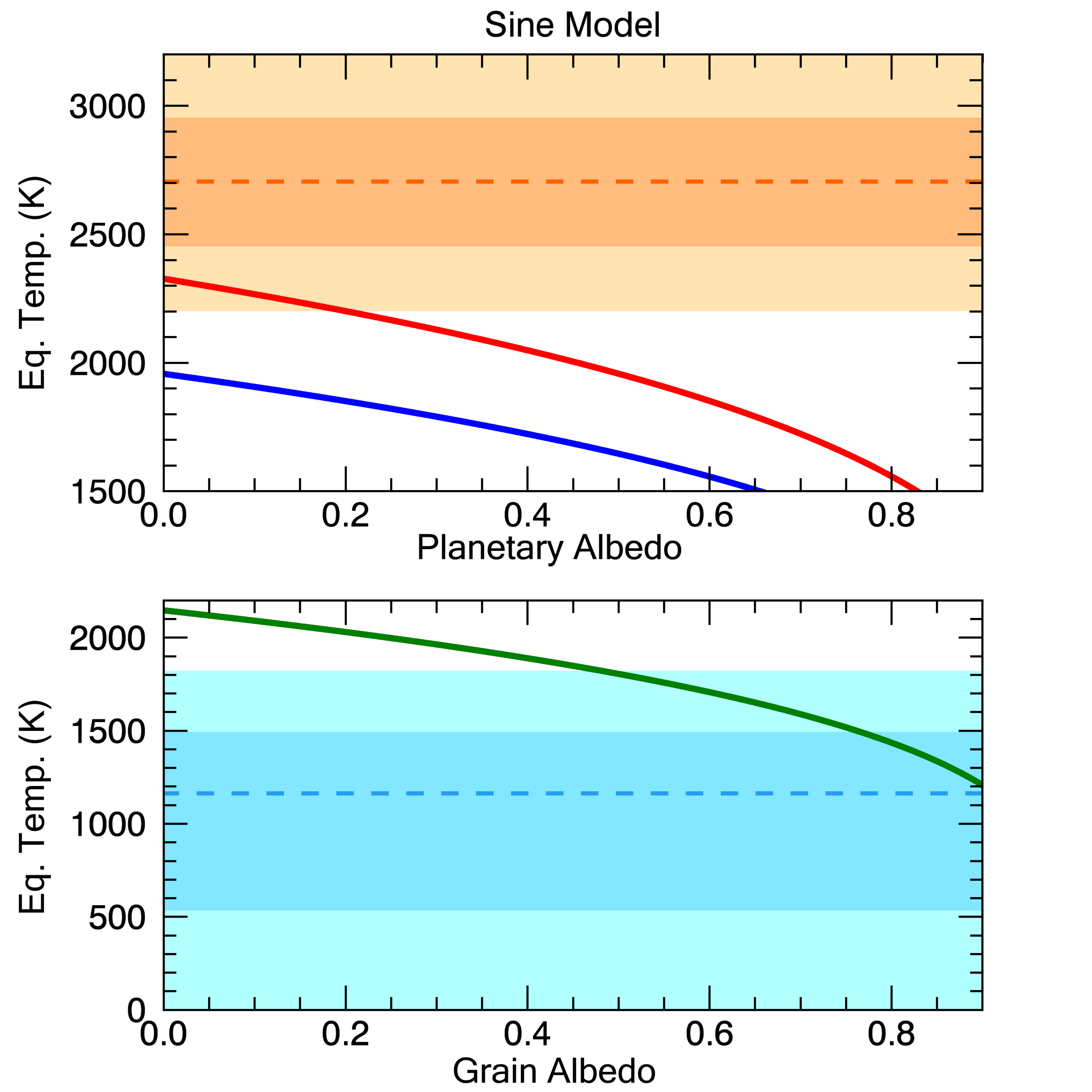} \\
\end{tabular}
\caption{Refractory particulate temperature calculations for our four variability models. The top panel in each shows planetary equilibrium temperatures for efficient (blue) and inefficient (red) cooling cases over a range of planetary albedos. The 2$\sigma$ range of $T_{b,max}$ for each model is shown in orange. The bottom panel shows the equilibrium temperature of refractory grains (green) at a height of 100 km above the planet's surface, due to radiation from the planet (at $T_{b,max}$) and the star. The 2$\sigma$ range of $T_{b,min}$ is shown in blue. For all four models, we find that $T_{b,max}$ can be achieved by a dark planet with inefficient heat redistribution, and $T_{b,min}$ could be achieved with reflective grains obscuring the hot planetary surface. Values for $T_{b,max}$ and $T_{b,min}$ are given in Table \ref{tab:brightnesstemps}.}
    \label{fig:btemps}
\end{figure*}

Next, for each model, we calculated the equilibrium temperature of refractory particulates at a height of 100 km above the planet's surface due to both solar and planetary radiation over a range of grain albedos. In each case, we assumed that the planet is radiating at its $T_{b,max}$ temperature. Our assumption of the height of the grain layer above the surface is about twice the maximum theoretical height of volcanic plumes on earth \citep[][]{Wilson1978}, which we feel is not unrealistic for a planet of $\sim$2R$_\oplus$ that may experience intense tidal heating. In practice, however, we found that our results for this calculation are not strongly dependent on the location of the layer of grains. The results of these calculations are shown in Figure \ref{fig:btemps}. The top panel shows the equilibrium temperature for efficient and inefficient cooling models compared to $T_{b,max}$ of the model, while the bottom shows the temperature of grains compared to $T_{b,min}$.  

While results vary for the different models, they all require a dark planetary surface (albedo $<$ 0.5) and grains that are reflective (albedo $>$ 0.4) at visible wavelengths where the star releases most of its energy, in order to be consistent with the 2$\sigma$ ranges of $T_{b,max}$ and $T_{b,min}$ based on planetary equilibrium temperature alone. They also require the inefficient cooling case to agree with $T_{b,max}$, which may be indicative of a lack of an atmosphere for efficient heat redistribution.

However, it is possible that the planet's thermal radiation budget is not limited to incoming flux from its star. If the four other planets in the system \citep[][]{Fischer2008} help maintain even a slightly eccentric orbit, 55 Cnc \textit{e} may dissipate a significant amount of radiation in the form of tidal heating. \citet{Demory2016b} examined tidal dissipation as a possible explanation for a 3100 K region in their longitudinal temperature map of 55 Cnc \textit{e}, using the Mercury-T code of \citet{Bolmont2015}. They found that tidal heat fluxes ranging from $10^{-3}$ to $10^6$ $W m^{-2}$ are achievable for the planet, corresponding to temperature increases of a few Kelvin to $\sim$2000 K. While the amount of tidal heating is largely unconstrained by current observations, it could represent a significant fraction of the planet's thermal radiation budget. Coupled with a low planetary albedo, tidal heating could significantly increase the surface temperature of the planet, which in turn could allow $T_{b,max}$ to be consistent with the efficient cooling model.

\citet{Budaj2015} tabulated the opacities and equilibrium temperatures of several species of dust grains that may be present in exoplanet atmospheres over a wavelength range from 0.2 to 500 $\mu$m and grain radii from 0.01 to 100 $\mu$m. Extracting their opacity results at 4.5 $\mu$m from their tables\footnote[2]{www.ta3.sk/$\sim$budaj/dust/}, we calculated the single scattering albedo (Eq. 23 from their paper) for all grain sizes from all species they examined. We used their temperature tables for a star of $T_{eff}$ = 5000 K, and solid angle of 0.01 and 0.03 steradians (55 Cnc subtends a solid angle of $\sim$ 0.02 sr in 55 Cnc \textit{e}'s sky). 

Cross-referencing these results, we identified the composition and size of grains that have the requisite albedos and equilibrium temperatures to match our brightness temperatures. These include alumina, olivine, and pyroxene, with grain radii ranging from around 0.01 to 0.6 $\mu$m. If the planet regularly condenses or outgasses these materials in an opaque cloud layer, its composition could achieve the appropriate albedo and equilibrium temperature to match $T_{b,min}$ from our models.

\citet{Mahapatra2017} applied a kinetic, non-equilibrium cloud formation model to the atmosphere of 55 Cnc \textit{e} using several different abundance cases. Using D16's derived T-P profile, they found that 55 Cnc \textit{e} can support a highly opaque cloud layer consisting of mainly Mg-silicates and Si-oxides. The local gas temperature of this cloud layer (see panel 1 of Fig. 10 in their paper) ranges from about 850-950 K as a function of pressure, which is within the 1$\sigma$ errors of $T_{b,min}$ for our year-to-year, slope, and sine models.

As a caveat, our model would require significant portions of the planetary surface to be blocked by a layer of cooler refractory grains, depending on the local structure of planetary surface temperature and the temperature of occulting material. Determining whether or not the maintenance of such a large layer of condensed or ejected material is feasible is beyond the scope of this simple calculation. 

\subsection{Future Work}
As we've discussed, more closely-spaced observations of 55 Cnc \textit{e}'s eclipse will be needed to constrain the timescale of thermal emission variability on the planet. The sine model created in Section \ref{subsec:modelcalcs} represents the short end of this timescale, with a period of $\sim$35 days. This model can be ruled in or out with Nyquist sampling of the planet's eclipse depth at 4.5 $\mu$m, by taking observations spaced at most 17.5 days apart over multiple periods. Further observations should place a stronger constraint on the magnitude of maximum/minimum brightness temperatures, which in turn may help discriminate between different absorbing species acting as an opaque cloud layer. 

Future space telescope missions may also allow us to better constrain the possibility of magma oceans and volcanism on highly irradiated rocky worlds such as 55 Cnc \textit{e}. As \citet{Kaltenegger2010} discuss, it may be possible to detect features of geochemical cycles indicative of volcanic activity on super-Earths through spectroscopy with the \textit{James Webb Space Telescope} \citep[][]{Gardner2006}. 

Unfortunately, 55 Cnc's brightness (mag. 4.0 in K$_s$ band, \citet{Skrutskie2006}) will saturate many observing modes of JWST. Recent performance tests by \citet{Greene2016}, however, indicate that spectroscopic observations of the system will be achievable using NIRCam's F444W filter. Tests performed with PandExo \citep{Batalha2017}, a tool that simulates observations with \textit{JWST}, confirm that observations of 55 Cnc are possible with NIRCam without saturating. The F444W filter, which offers wavelength coverage from $\sim$3.9-5.0 $\mu$m, will not likely constrain the presence of silicate materials, which have the strongest spectral features between 8 and 12 $\mu$m \citep[][]{Hu2012}. Nevertheless, observations with NIRCam will be useful for further probing the variability of 55 Cnc \textit{e} and for searching for features indicative of the presence of an atmosphere.

Additionally, a mode proposed by \citet{Schlawin2017} using NIRCAM's Dispersed Hartmann Sensor (DHS) may be able to take observations of comparably bright objects, as it spreads light over 10 separate spectra. Whether this proposed mode will be approved, and whether or not it will be able to image 55 Cnc without saturating, remain to be seen.  

If selected to go forward, the 55 Cnc system may also be observable with the proposed \textit{ARIEL} mission \citep[see][]{Tinetti2016}, which will be designed to monitor the atmospheres of exoplanets, some of which will be in the super-Earth size range. Such a mission would also allow us to better constrain the presence or lack of an atmosphere on 55 Cnc \textit{e}.

\section{Conclusions}

In this work, we presented a reanalysis of five transit and eight eclipse observations of the super-Earth 55 Cnc \textit{e} taken in the 4.5 $\mu$m channel of the \textit{Spitzer Space Telescope}. We made this reanalysis using the pixel-level decorrelation (PLD) framework of \citet{Deming2015}. The eclipse observations were found by \citet{Demory2016a} to exhibit a significant amount of variability on the timescale of a year, which D16 attributed to volcanic outgassing. We find that the transit depth of the planet is consistent with a constant value, so we cannot confirm the possible transit depth variability reported in D16.

Using orbital parameters derived from a Markov Chain Monte Carlo analysis of the transit observations, we performed our PLD regression analysis on the eclipse data, and explored a number of parameters that can impact the retrieved eclipse models, such as bin size, number of basis pixels used, and amount of data fit. Creating models to fit our eclipse results, we find that we can confidently conclude that the planet's eclipse depth \textit{is} variable. However, future observations are needed to determine the timescale of eclipse depth variability. We fit five different models to our eclipse results, and calculating brightness temperatures for each of our models, investigated if the planetary equilibrium temperature and obscuring volcanic grains can reproduce $T_{b,max}$ and $T_{b,min}$. We find that our models require a dark planet with inefficient cooling and reflective ejecta to be consistent with our brightness temperatures, if stellar insolation is the only source of heat on 55 Cnc \textit{e}. 

\acknowledgments

We thank Heather Knutson for her valued input on fixing orbital parameters in our MCMC procedure. We also thank Prabal Saxena for his guidance on tidally heated exoplanets. Finally, we thank our anonymous referees for their valuable input for  revisions. This work is based on archival data obtained with the Spitzer Space Telescope, which is operated by the Jet Propulsion Laboratory, California Institute of Technology under a contract with NASA. Support for this work was provided by NASA. We gratefully acknowledge support from NASA grant NNX16AF34G.



Facilities: \facility{Spitzer(IRAC)}.

\clearpage

\end{document}